\documentclass[sigconf,screen]{acmart}
\newcommand{{\proj}}[0]{Collab}
\usepackage{algorithm}
\usepackage{pifont}
\usepackage{algpseudocode}
\usepackage{graphicx}
\usepackage{tikz}
\usepackage{tabularx}
\usetikzlibrary{trees}
\usepackage{subfig}
\usepackage{diagbox}
\usepackage[table]{xcolor} 

\definecolor{lightBarBlue}{HTML}{ADC8E6}  
\definecolor{mediumBarBlue}{HTML}{77A2C9} 
\definecolor{darkBarBlue}{HTML}{4A7AA1}   

\AtBeginDocument{%
  \providecommand\BibTeX{{%
    \normalfont B\kern-0.5em{\scshape i\kern-0.25em b}\kern-0.8em\TeX}}}

\copyrightyear{2026}
\acmYear{2026}
\setcopyright{cc}
\setcctype{by}
\acmConference[CHI '26]{Proceedings of the 2026 CHI Conference on Human Factors in Computing Systems}{April 13--17, 2026}{Barcelona, Spain}
\acmBooktitle{Proceedings of the 2026 CHI Conference on Human Factors in Computing Systems (CHI '26), April 13--17, 2026, Barcelona, Spain}
\acmPrice{}
\acmDOI{10.1145/3772318.3790501}
\acmISBN{979-8-4007-2278-3/2026/04}

\begin{document}

\title[Collab: Fostering Critical Identification of\\ Deepfake Videos on Social Media via Synergistic Annotation]{Collab: Fostering Critical Identification of Deepfake Videos on Social Media via Synergistic Annotation}

\author{Shuning Zhang}
\orcid{0000-0002-4145-117X}
\affiliation{%
  \institution{Tsinghua University}
  \city{Beijing}
  \country{China}
}
\email{zsn23@mails.tsinghua.edu.cn}

\author{Linzhi Wang}
\orcid{0009-0000-3860-7367}
\affiliation{%
  \institution{Tsinghua University}
  \city{Beijing}
  \country{China}
}
\email{wang-lz22@mails.tsinghua.edu.cn}

\author{Shixuan Li}
\orcid{0009-0008-6828-6347}
\affiliation{%
  \institution{Tsinghua University}
  \city{Beijing}
  \country{China}
}
\email{li-sx24@mails.tsinghua.edu.cn}

\author{Yuanyuan Wu}
\orcid{0009-0004-2633-9691}
\affiliation{%
  \institution{Shanghai Jiaotong University}
  \city{Shanghai}
  \country{China}
}
\email{buddy.yuan@sjtu.edu.cn}

\author{Yuwei Chuai}
\orcid{0000-0001-6181-7311}
\affiliation{
  \institution{University of Luxembourg}
  \city{Luxembourg}
  \country{Luxembourg}
  }
\email{yuwei.chuai@uni.lu}

\author{Luoxi Chen}
\orcid{0009-0007-8213-0683}
\affiliation{%
  \institution{Tsinghua University}
  \city{Beijing}
  \country{China}
}
\email{chenlx22@mails.tsinghua.edu.cn}

\author{Xin Yi}
\orcid{0000-0001-8041-7962}
\authornote{Corresponding author.}
\affiliation{
    \institution{Tsinghua University}
    \city{Beijing}
    \country{China}
}
\email{yixin@tsinghua.edu.cn}

\author{Hewu Li}
\orcid{0000-0002-6331-6542}
\affiliation{
    \institution{Tsinghua University}
    \city{Beijing}
    \country{China}
}
\affiliation{
    \institution{Zhongguancun Laboratory}
    \city{Beijing}
    \country{China}
}
\email{lihewu@cernet.edu.cn}

\renewcommand{\shortauthors}{Zhang, et al.}


\begin{abstract}

Identifying deepfake videos on social media platforms is challenged by dynamic spatio-temporal artifacts and inadequate user tools. This hinders both critical viewing by users and scalable moderation on platforms. Here, we present Collab, a web plugin enabling users to collaboratively annotate deepfake videos. Collab integrates three key components: (i) an intuitive interface for spatio-temporal labeling where users provide confidence scores and rationales, facilitating detailed input even from non-experts, (ii) a novel confidence-weighted spatio-temporal Intersection-over-Union (IoU) algorithm to aggregate diverse user annotations into accurate aggregations, and (iii) a hierarchical demonstration strategy presenting aggregated results to guide attention toward contentious regions and foster critical evaluation. A seven-day online study (N$=$90), where participants annotated suspicious videos when viewing an online experimental platforms, compared Collab against two conditions without aggregation or demonstration respectively. Collab significantly improved identification accuracy and enhanced reflection compared to non-demonstration condition, while outperforming non-aggregation condition for its novelty and effectiveness.
\end{abstract}



\begin{teaserfigure}
    \centering
    \includegraphics[width=1\linewidth]{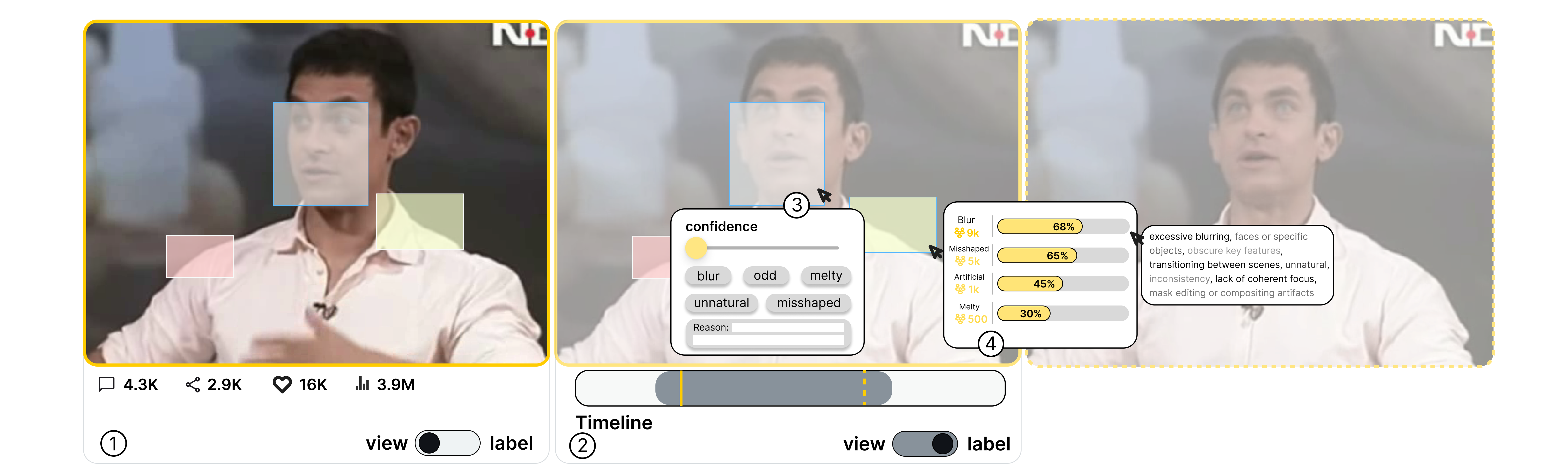}
    \caption{The \proj{} interface for collaborative deepfake videos' annotation. \textcircled{1} On videos potentially containing deepfake artifacts, \proj{} overlays existing community annotations (semi-transparent regions). Users can toggle into labeling mode to contribute. \textcircled{2} Users then draw rectangular spatio-temporal regions directly onto the video player. \textcircled{3} Selection completion reveals an annotation panel for inputting label, confidence, refined time range (via timeline), and optional rationale. \textcircled{4} Peer annotations are viewable; the panel (\textcircled{3}) shows aggregated metrics (confidence, count), while individual rationales are hover-revealed to mitigate social conformity bias. Clicking a label accesses detailed aggregates.}
    \label{fig:teaserfigure}
\end{teaserfigure}


\maketitle

\section{Introduction}

The proliferation of online misinformation poses a persistent challenge to informed public discourse~\cite{vosoughi2018spread,wang2019systematic,argentino2021qanon,bengali2019whatsapp,dixit2018whatsapp,thomas2022doctors}. This issue is further exacerbated by intentional disinformation, particularly when advanced manipulative techniques, such as deepfakes, are employed~\cite{tahir2021seeing,zhou2025effect}. In response, social media ecosystems have predominantly relied on centralized, platform-controlled governance mechanisms that integrate automated detection~\cite{alrashidi2022review}, large-scale human moderation~\cite{arsht2018human}, and third-party fact-checking partnerships~\cite{belair2023knowledge}. However, these platform-centric strategies face significant limitations. Automated tools often exhibit brittleness against novel manipulation techniques, while human-led efforts struggle with scalability bottlenecks, substantial costs, and consistency issues~\cite{arora2023detecting, arsht2018human}. Furthermore, the operational opacity of these systems can limit user agency and foster distrust~\cite{gorwa2020algorithmic}.

These limitations fueled a trend towards community-driven crowd-sourced approaches like $\mathbb{X}$'s Community Notes~\cite{communitynotes2022diversity}. Indeed, studies suggest that empowering users can foster critical thinking and reduce the spread of manipulated content~\cite{jahanbakhsh2021exploring,pennycook2021shifting,pennycook2020fighting,jahanbakhsh2022leveraging,jahanbakhsh2024browser}. However, sophisticated video deepfakes introduces unique and under-explored challenges: \textit{existing user-centric tools are often ill-equipped for the spatio-temporal nature of video, lacking interfaces for the fine-grained annotation required to pinpoint manipulations precisely}. Furthermore, translating diverse and potentially conflicting user annotations into a reliable collective judgment requires robust aggregation mechanisms. It also remains a critical challenge to present these aggregated insights to constructively enhance user assessment while mitigating the risks of detrimental social influence~\cite{ismailoglu2022aggregating}. To address these challenges, we ask the following research questions (RQs):

\noindent $\bullet$ RQ1: How can theories of Collective Intelligence (CI) and Social Influence (SI) inform a design framework for collaborative deepfake identification?

\noindent $\bullet$ RQ2: How can such a framework be instantiated into a system with effective mechanisms for spatio-temporal annotation, aggregation and demonstration?

\noindent $\bullet$ RQ3: To what extent does the design of the annotation tool enable users to accurately and efficiently annotate deepfake videos?

To address these questions, we first developed a theoretically-grounded design framework specifically tailored for collaborative deepfake video identification. This framework provides a coherent structure by explicitly modeling and organizing the interconnected processes of labeling, aggregation, and demonstration (Figure~\ref{fig:design_dimension}), recognizing their inherent interdependencies where aggregation bridges labeling and demonstration, and demonstration feeds back into labeling. We further leveraged Collective Intelligence (CI)~\cite{wolpert1999introduction} and Social Influence (SI)~\cite{cialdini2004social} theories to guide the framework. With these grounds, we operationalized key principles: the labeling and aggregation components are shaped by CI, emphasizing \textbf{diversity} (L1, A1), \textbf{independence} (L2, A2) and \textbf{motivation} (L3, A3) through specific design choices. The demonstration component is structured by SI, managing potential \textbf{Normative Influence}, \textbf{Informational Influence} and \textbf{Social Comparison} through controlled information \textbf{visibility} (D1), \textbf{density} (D2) and \textbf{social cues} (D3). 

In accordance with the framework and guidelines, we then designed and implemented \proj{}, an annotation tool on social media platform, architected around the interconnected processes of annotation, aggregation and demonstration. \textbf{For annotation}, users employ a direct manipulation interface to define spatio-temporal regions on videos, mirroring familiar paradigms like bullet-screen commenting (Figure~\ref{fig:teaserfigure}). This process involves assigning predefined or custom labels, specifying a confidence level, and optionally providing textual justifications, thereby facilitating nuanced quality assessment. Subsequently, \textbf{for aggregation}, a novel aggregation mechanism processes these individual annotations using a confidence-weighted 3D Intersection-over-Union (IoU)-based aggregation algorithm. This method iteratively merges annotations based on spatio-temporal overlap and confidence scores, generating consolidated regions and determining predominant labels weighted by both instance confidence and user reliability metrics. \textbf{For demonstration,} the resulting aggregated insights are then presented back to users through a layered visualization strategy. Color-coded overlays provide initial cues, with further details like dominant labels, aggregated confidence, and summarized rationales revealed hierarchically upon user interaction. This demonstration approach is carefully structured to convey collective intelligence effectively while minimizing potential negative social influence by strategically withholding raw peer data.

We evaluated \proj{} by conducting a 7-day online study (N=90) following established practices~\cite{jahanbakhsh2022leveraging,jahanbakhsh2024browser}. Using a custom experimental social platform, we compared \proj{} against two alternative conditions: one without the aggregation mechanism and another without the demonstration of peer labels. \proj{} achieved an overall accuracy of 88.1\%, which was significantly higher than the \textit{No Agg} condition (79.7\%). \proj{} also significantly outperformed the \textit{No Label} and \textit{No Agg} condition in subjective ratings such as \textit{Reflection}, \textit{Rational} judgment, \textit{Smoothness}, \textit{Novel} and \textit{Effective}. Users praised \proj{} for its critical thinking and reflective assessments, which avoided negative social influence such as conformity bias~\cite{moscovici1972social}. Together, the contributions of this work are threefold:

\noindent $\bullet$ We proposed a theoretically-grounded design framework for collaborative deepfake video identification, which structures the interconnected processes of labeling, aggregation and demonstration.

\noindent $\bullet$ We presented Collab, a collaborative labeling technique incorporating an intuitive spatio-temporal annotation interface, a confidence-weighted 3D label-merging algorithm for aggregation, and a hierarchical visualization strategy for demonstration.

\noindent $\bullet$ A 7-day online evaluation (N=90) demonstrated \proj{}'s effectiveness in significantly improving deepfake video identification accuracy and enhancing critical annotation compared to alternatives.

\section{Background \& Related Work}

We review foundational literature concerning deepfake videos, its social influence and collaborative countermeasures. We examine diverse identification and mitigation strategies, alongside user interface design principles for collaborative annotation, with a focus on video-specific challenges.

\subsection{Deepfake Detection For Machines and Humans}

The proliferation of sophisticated deepfakes led to two primary lines of research: automated technical methods and human-centered systems. Automated detection remains a significant challenge. While numerous deepfake datasets have been developed, such as DFDC~\cite{dolhansky2020deepfake} and DF-W~\cite{pu2021deepfake}, they are often criticized for being unbalanced~\cite{layton2024sok}. Besides, current detectors struggle to generalize and often fail to identify high-quality, ``in-the-wild'' deepfakes~\cite{pu2021deepfake,yan2023deepfakebench}. Furthermore, these automated systems are vulnerable to adversarial attacks, where carefully crafted, imperceptible perturbations can cause a detector to misclassify a fake video as authentic~\cite{hussain2021adversarial,shahriyar2022evaluating}. 

Concurrently, research on human perception reveals that unaided judgment is often unreliable~\cite{tahir2021seeing}, with detection accuracy hovering just above chance levels~\cite{diel2024human}. Synthesis engines advanced past the ``uncanny valley'', creating faces that are not only indistinguishable from real ones but are sometimes perceived as more trustworthy~\cite{nightingale2022ai}, leaving users vulnerable to deception~\cite{mink2022deepphish}. A promising direction for mitigating these individual errors is to aggregate diverse judgments. This principle has been demonstrated to be effective in both human-AI collaboration~\cite{groh2022deepfake} and evident in crowdsourced annotations on static deepfake images~\cite{joslin2024double}. However, this paradigm has not yet been adapted to the unique spatio-temporal complexities of video. Building on this foundation, we designed \proj{}, a collaborative technique harnessing crowdsourced intelligence for deepfake videos' identification.

\subsection{Sociotechnical Dynamics and Human-Centered Approaches}

Understanding the social context in which deepfakes are encountered is crucial for developing effective interventions. Perceptions of deepfakes vary across communities. Some online forums actively foster pro-deepfake cultures, building communities and marketplaces for creating and sharing synthetic media~\cite{gamage2022deepfakes}. Conversely, in many low-resource communities, awareness of sophisticated digital manipulation is low, and users often equate ``fake'' with factually inaccurate content rather than doctored media~\cite{shahid2022matches}. Even among trained professionals like journalists, over-reliance on imperfect detection tools can sometimes compromise their verification processes~\cite{sohrawardi2024dungeons}. Moreover, human moderation is susceptible to cognitive biases, where judgments can be influenced by racial or gender stereotypes~\cite{mink2024s}.

These challenges have motivated a shift toward human-centered and collaborative systems. Research shows that human performance can be significantly improved through targeted interventions such as feedback and training programs~\cite{diel2024human, tahir2021seeing}. New tools are being designed with analysts' specific requirements in mind to support interactive investigation~\cite{wu2025understanding}. Leveraging unique human perceptual cues, such as distinct eye movement patterns when viewing real versus fake videos, aids in forgery localization~\cite{gupta2020eyes}. Critically, crowdsourcing has emerged as a promising paradigm. For instance, Joslin et al.~\cite{joslin2024double} demonstrated that leveraging crowdsourced annotations of facial regions can effectively train more robust detectors for synthesized images. This body of work underscores the potential for systems that harness collective human intelligence, providing a strong rationale for developing structured, collaborative annotation frameworks to address the deepfake challenge.

\subsection{User Interface Design for Collaborative Annotating}

Collaborative annotation techniques, often leveraged in crowdsourcing, have been examined across theoretical, application and technical dimensions. \textbf{Theoretical-wise,} Stureborg et al.~\cite{stureborg2023interface} studied whether and how concept hierarchies can inform the design of annotation interfaces to improve labeling quality and efficiency. Hartwig et al.~\cite{hartwig2024landscape} surveyed user-centered misinformation interventions and proposed taxonomies of collaborative annotations. Most studies examine individual components for single-user annotations rather than developing video-oriented theories.

\textbf{Application-wise,} Bhuiyan et al.~\cite{bhuiyan2023newscomp} proposed comparative news annotation, which annotated similarities and differences between pairs of articles. Wood et al.~\cite{wood2018rethinking} built a mobile app ``Newsr'' to support co-annotation in the form of graffiti, on online news articles. These implementations show collaborative paradigms but lack the specific focus on video annotation or governance.

\textbf{Technical-wise, researchers focused on human-AI collaborative or multi-user collaborative techniques.} Park et al.~\cite{park2024leveraging} discussed the potential to leverage LLMs as an interactive research tools to facilitate collaboration between human coders and AI to effectively annotate online risk data at scale. Shabani et al.~\cite{shabani2021sams} introduced a human-in-the-loop approach to combat the sharing of misinformation, which leverages the fact-checking skills of humans by providing feedback on news stories about the source, author, message and spelling. Jahanbakhsh et al.~\cite{jahanbakhsh2022leveraging} designed a prototype social media platform that implement these user affordance as structured inputs to directly impact how and whether posts are shown. They~\cite{jahanbakhsh2024browser} further designed an in-browser signaling prototypes to support collaborative trusted annotations. However, all the previous platforms did not consider the annotation on rich multimodal data such as videos. These modalities brought inherent challenges for annotation, which we aimed to cope with.

Recent work explores embedding inline credibility signals into video player, such as provenance, source metadata, and concise contextual evidence, to aid viewer trustworthiness judgments~\cite{hughes2024viblio}. Concurrently, at the platform level, systems like $\mathbb{X}$'s Community Notes show how peer-authored contextual annotations and community explanations can surface crowd judgments and signal contested content~\cite{communitynotes2022diversity}. Furthermore, the widely recognized C2PA specification establishes standards for certifying content provenance and detecting tampering~\cite{rosenthol2022c2pa, feng2023examining}. While C2PA provides a crucial foundation in file integrity, its utility depends on a trustworthy content production. Our work complements this by contributing crowd-sourced detection methods applicable even to deepfakes generated by non-compliant or open-source models. Similarly, platform interventions like Community Notes and YouTube Shorts' contextual notes~\cite{youtube2024new} primarily rely on holistic, post-level text signaling. Our work improves these efforts by investigating fine-grained spatio-temporal marking specifically tailored for complex video content.

\section{Design and Implementation of Collab}\label{sec:design_implementation}

\subsection{Design Guidelines}\label{sec:goal}

\subsubsection{Overall Design Goals}

\proj{} is designed to enhance effectiveness, user experience and critical thinking in deepfake video identification. To tackle the unique challenges of dynamic video content and the socio-cognitive demands of collaborative sensemaking, our design framework integrates core principles from Collective Intelligence (CI)~\cite{wolpert1999introduction} and Social Influence~\cite{cialdini2004social} theories. These theories directly inform our high-level design goals, which are formulated to address the interplay between human perception, cognitive constraints, and social factors inherent in video annotation:

\begin{figure*}[h]
    \centering
    \includegraphics[width=0.8\textwidth]{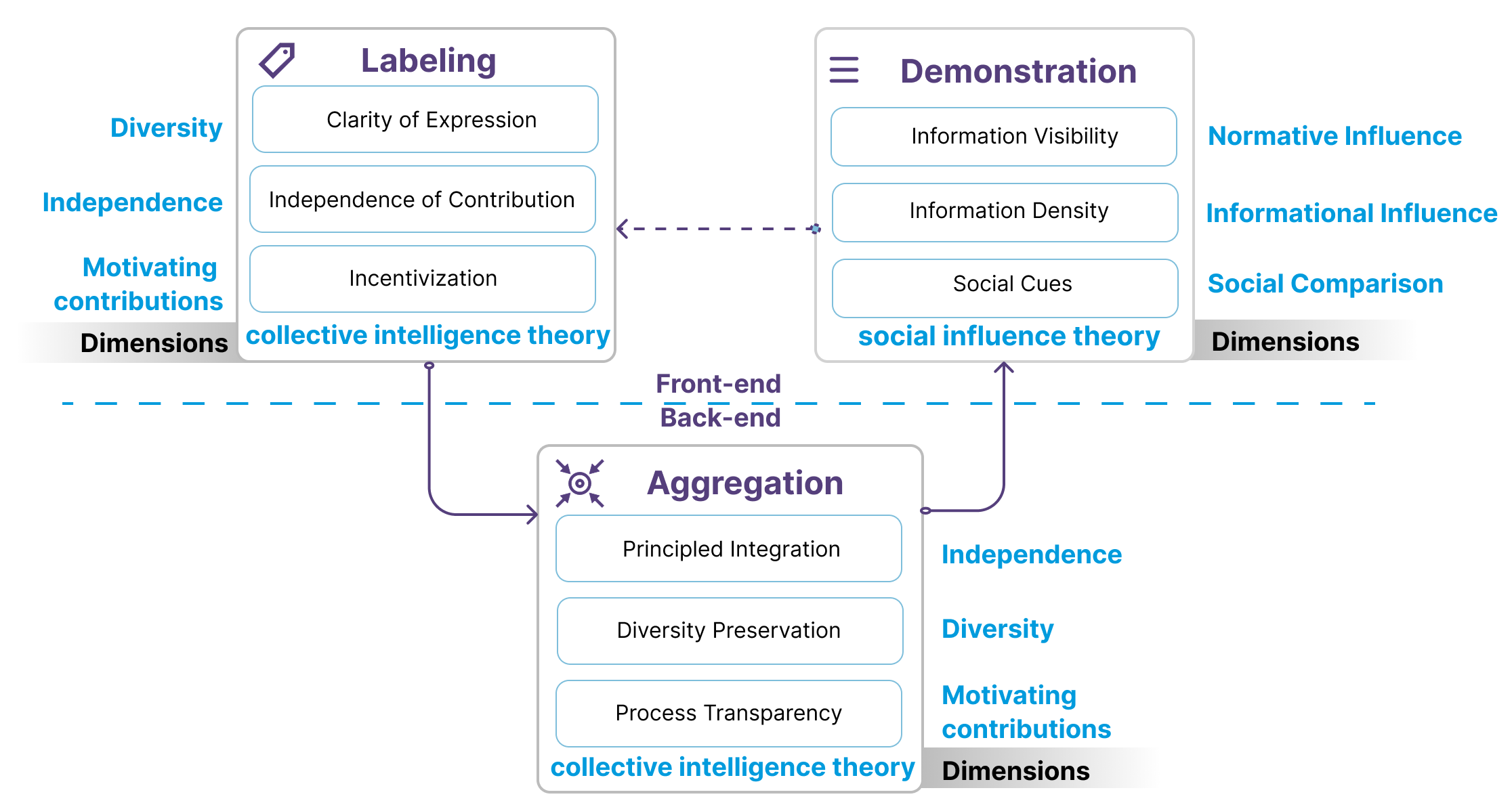}
    \caption{The design dimensions of Collab.}
    \label{fig:design_dimension}
\end{figure*}

\noindent $\bullet$ \textbf{(G1) Leverage Human Temporal Reasoning:} Harness users' unique capacity to interpret dynamic, sequential information inherent in video, ensuring annotations capture temporal context and nuance beyond static frames. This serves as a crucial foundation for enhancing \textbf{CI} and facilitating aggregation processes.

\noindent $\bullet$ \textbf{(G2) Mitigate Bias and Modulate Social Influence:} Structure interactions to minimize detrimental cognitive biases (e.g., anchoring~\cite{furnham2011literature}, confirmation bias~\cite{klayman1995varieties}) while leveraging constructive \textbf{SI} (e.g., exposure to diverse perspectives) to enhance collective accuracy. This helps increase the accuracy of \textbf{CI} outcomes.

\noindent $\bullet$ \textbf{(G3) Optimize Annotation Workflow:} Streamline the workflow to ensure usability and minimize cognitive load during interaction with complex video data. This facilitates readily contributed individual inputs, providing the foundation for \textbf{CI} and \textbf{SI}.

\noindent $\bullet$ \textbf{(G4) Foster Critical Engagement:} Encourage users to reflect on their own judgments and the inputs of others, promoting critical engagement with the content and the annotation task itself. This objective leverages \textbf{SI} constructively, fostering the critical engagement of the \textbf{CI} process.

To operationalize these goals, we structure the user interaction flow into labeling, aggregation and demonstration processes, with specific guidelines tailored for each process.

\subsubsection{Design Guidelines for Each Process}

These aforementioned goals guide the instantiation of specific dimensions across the three core processes: labeling, aggregation and demonstration, explicitly drawing upon CI and SI principles~\cite{luo2023learning,huang2019neural}.

\paragraph{Labeling} CI theory~\cite{wolpert1999introduction} emphasizes the necessity of diversity, independence, and the motivation to generate high-quality contributions. We propose three guidelines corresponding to these dimensions. First, the interface should ensure \textbf{Clarity of Expression (L1)} by enabling users to clearly articulate their interpretations, including labels, confidence labels, and rationales. This requires structured input mechanisms~\cite{lacerda2014does}, such as predefined categories and rationale prompts, while retaining the flexibility to highlight specific regions or times (G1,G4). Second, to prevent negative conformity effects~\cite{rosander2012conformity,park2010formity,bernheim1994theory}, the system should enforce \textbf{Independence of Contribution (L2)}. Initial annotations should be made independently before exposure to others' inputs, aligning with CI principles that require decentralized knowledge aggregation (G2). Third, \textbf{Incentivization (L3)} mechanisms are essential to motivate users and enhance collective input (G3,G4). This can be achieved by encouraging high-quality, thoughtful contributions through feedback on rationale thoroughness or confidence calibration~\cite{kuang2019spillover,claussen2013effects}.

\paragraph{Aggregation} CI theory~\cite{wolpert1999introduction} also guides how individual annotations are combined into collective representations. We propose three guidelines regarding independence, diversity and transparency to balance objective synthesis with comprehensive representation. We employ \textbf{Principled Integration (A1)}, using aggregation methods such as confidence weighting~\cite{wang2011aggregating}, and spatial IoU calculations~\cite{cheng2021boundary,zhou2019iou} to quantitatively consolidate multi-source data into robust collective judgments. Simultaneously, the system prioritizes \textbf{Diversity Preservation (A2)} to avoid premature convergence. Instead of forcing a single consensus, aggregation should surface the full range of perspectives and rationales, such as through clustering valid differing interpretations, to expose users to the diverse viewpoints crucial for critical thinking (G2,G4). Finally, we ensure \textbf{Process Transparency (A3)} so that users understand how aggregated results are formed from individual contributions. Visualizing label distributions or confidence levels fosters trust and enables informed interpretation (G4).

\paragraph{Demonstration}
The demonstration stage involves presenting the aggregated annotations back to users. Drawing from SI theory~\cite{cialdini2004social}, the design should carefully consider how information presentation influences user perception and subsequent actions. Specifically, we manage three dominant types of influence: normative, informational, and social comparison. First, we control \textbf{Information Visibility (D1)}, as the prominence and timing of displaying collective annotations impact normative influence~\cite{cialdini2004social}. While high visibility may increase conformity and decision speed, it risks suppressing independent judgment (G2). Second, we calibrate \textbf{Information Density (D2)}, because the amount and complexity of aggregated information affect cognitive load and informational influence~\cite{cialdini2004social}. High density might increase acceptance of others' views but can lead to cognitive overload (G2,G3). Third, the system utilizes explicit \textbf{Social Cues (D3)}, such as the number of agreements~\cite{mehu2014multimodal}, to trigger social comparison processes~\cite{suls2002social} that influence conformity and trust (G2).

\subsection{System Design}

Users opened the social media platform (e.g., $\mathbb{X}$, TikTok) to view the videos, with the regions of others' labels shown by default. They could toggle the button if they identified potential deepfake videos (Figure~\ref{fig:teaserfigure}). To create an annotation, they executed a draw-and-hold gesture on the screen. The temporal duration of this gesture defined the time range being marked, while the drawn rectangle defined the spatial region. They needed also to input a label (preset or customized by themselves), a confidence level and an optional reason. Upon their labeling, a timeline popped-up below the video player for users, showing the annotated time range. Users could hover upon the labeled regions to see the aggregated labels, confidence and reasons of these annotations. Through this process, users collaboratively labeled the videos with critical engagement, resulting in high accuracy after aggregation. We introduce the interaction design of \proj{} through three interconnected processes: annotation design, aggregation design, and label demonstration design, each guided by the specific design dimensions detailed in Section~\ref{sec:goal}.

\subsubsection{Annotation Design}

We designed the annotation tailored to the video modality, drawing inspiration from intuitive annotation methods such as progress bar~\cite{harrison2007rethinking} and interactive markers~\cite{bianco2015interactive}. We adopted metaphors familiar to users: video editing software~\cite{schoeffmann2015video} and bullet screen commenting systems\footnote{Commercial media platforms like Tiktok live-streaming and Bilibili have bullet screen comments.}~\cite{djamasbi2016social,wan2020online}. These paradigms help lower the learning curve (G4) in two key ways: 1) users' annotations are visualized as boxes on the screen and regions on the timeline below the video player, similar to those in video editing software (G4), 2) users need to select the label, indicate the confidence and reasons to annotate, a process similar to sending bullet-screen comments, which lowers the learning cost (L1,G3) and provides incentivization (L3).

\paragraph{Annotation Interface and Workflow}

To make the annotation process non-intrusive and intuitive (G3), \proj{} adopted an annotation paradigm that allows users to annotate while interacting with the video, mirroring common social media commenting~\cite{TikTokCommentsGuide} or video editing tool use~\cite{truong2016quickcut}. The interface provides users with the ability to:

\textbf{Toggle Annotation Mode:} The `Annotation Mode' enables users to edit annotations (add, change, delete), mirroring bullet screen control mechanisms~\cite{Li2016} and corresponding commercial implementations (e.g., Bilibili~\footnote{\url{https://www.bilibili.com/}}). An on/off toggle for annotation display aligns with users' established mental models. However, even with the toggle off, others' annotations remain visible. This design choice stimulate critical engagement and prompt users to assess video authenticity, thereby encouraging contribution.

\textbf{Highlight Spatio-Temporal Segments:} Users define video artifact segments by selecting rectangular regions. The time segments of these regions are then highlighted on the timeline below the video player. Users can modify time segments directly on the progress bar (e.g., dragging a slider or using start/end markers), during which the video would be automatically paused to facilitate marking the duration of the deepfake artifacts (G1,G3). The time-based annotations are represented as colored blocks on the progress bar, clearly showing the annotated ranges. This design aligns with the \textit{theories of spatial organization}~\cite{beyes2012spacing}. To maintain workflow continuity, we prevented the modification of spatial and temporal annotation boundaries after initial placement. This design choice is supported by pilot findings indicating that participants generally annotate deliberately, making subsequent boundary adjustments unnecessary. 

\textbf{Edit and Submit Annotations:} Upon marking a relevant segment, users are prompted to submit structured annotations (G4). The annotation protocol requires three components: the label of the deepfake video artifact, a confidence score expressed as a percentage (0-100\%), and an optional free-text justification. The selection and design of these components were guided by prior research~\cite{lee2015using} to enhance data precision (L1) while maintaining user comprehensibility (G3). We pre-defined the label taxonomy based on prior literature~\cite{joslin2024double}, \textbf{and also allowed custom labels if users think the artifacts do not belong to existing categories}. The pre-defined taxonomy include the following thirteen categories: \textit{blurry}, \textit{unnatural skin}, \textit{distorted}, \textit{strange texture}, \textit{strange shape}, \textit{strange skin folds}, \textit{irregular shape}, \textit{non-existent/unneeded object}, \textit{artificial}, \textit{mismatch}, \textit{melting}, \textit{molten metal}, and \textit{artificial material}. Providing these established categories, consistent with Joslin et al.~\cite{joslin2024double}, facilitates easy and potentially high-quality annotation. Regarding the confidence score, we utilize a continuous confidence scale rather than discrete bins (e.g., low/high). Computationally, it reduces quantization error for granular aggregation. Cognitively, it allows users to express probability without the friction of mapping uncertainty to coarse categories, thereby mitigating ceiling effects~\cite{hessling2004ceiling}.

\textbf{Assistance From Others' Labels}: To improve annotation accuracy and user engagement, users can view aggregated annotations from others as overlays (e.g., heatmap or markers) on the video (G2) (see Sections~\ref{sec:aggregation} and~\ref{sec:demonstration}) for processing and demonstration details). This design is informed by social influence theory~\cite{friedkin1998structural}, particularly informational influence~\cite{sussman2003informational}. We visualize the aggregated labels instead of each raw individual annotations (see Section~\ref{sec:demonstration}) to ease users' viewing efforts and constructively guide attention. We use different color gradients to separate different labeling regions, with confidence and multi-user agreement levels acting as criteria (detailed parameters set with a pilot study in Appendix~\ref{sec:color}):

\begin{itemize}
    \item Green: High-confidence, high-agreement annotations.
    \item Orange: Medium-confidence, some disagreement.
    \item Red: Low-confidence, uncertain or contested areas.
\end{itemize}

The red color indicates regions needing further clarification, whereas green color indicates those regions where the consensus is largely resolved. To mitigate bias, we refrained from using others' labels as anchors to guide users' selection (Figure~\ref{fig:collab_design} \textcircled{1}, \textcircled{2}, \textcircled{4}, \textcircled{6}, see Appendix~\ref{sec:annotation_comparison} for detailed comparisons). This strategy ensures that annotation process is relatively independent unless users proactively seek the help of others' labels (L2,G2). 

\begin{figure*}[!htbp]
    \centering
    \subfloat[Design choices of region boxes.]{
        \includegraphics[width=0.8\textwidth]{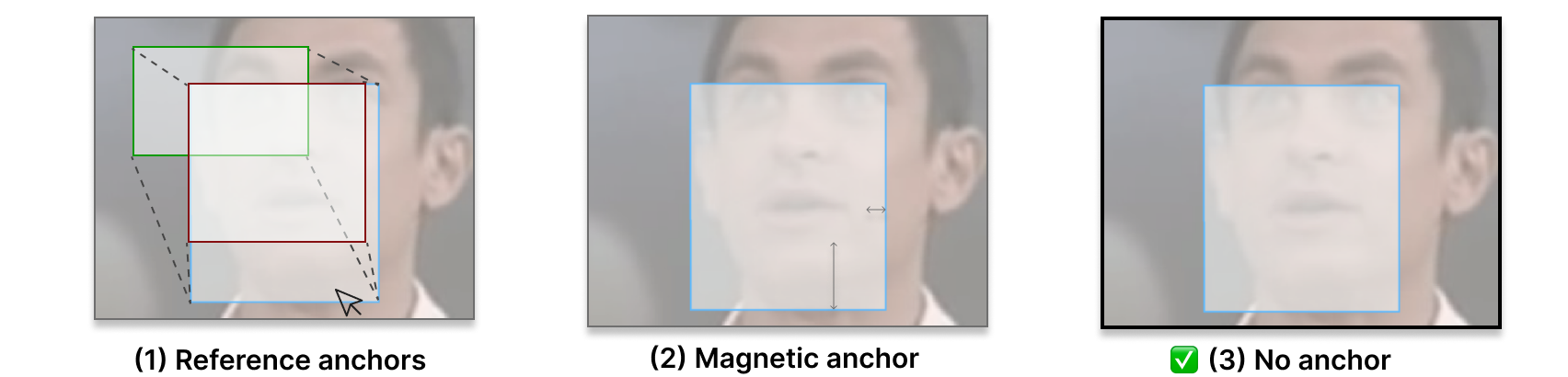}
        \label{fig:design_choice_box}
    }

    \subfloat[Design choices of labels.]{
        \includegraphics[width=0.8\textwidth]{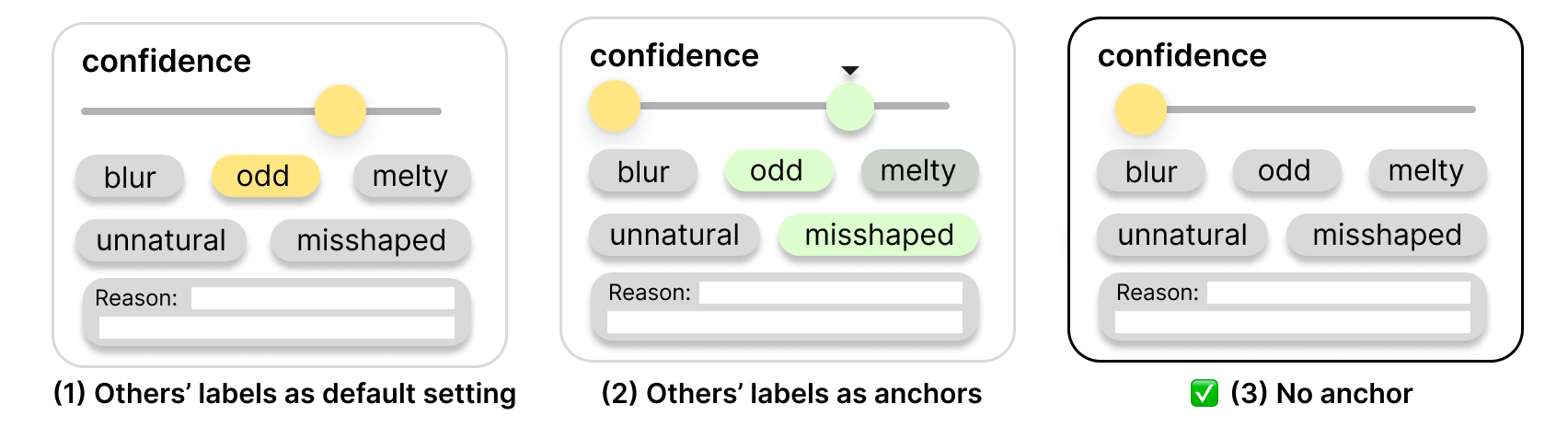}
        \label{fig:design_choice_label}
    }
    \caption{The design candidates of \proj{}, where the choices selected by \proj{} are highlighted with green marks.}
    \label{fig:collab_design}
\end{figure*}

\subsubsection{Aggregation Design}\label{sec:aggregation}

To effectively utilize users' collective intelligence, we propose a novel \textbf{Confidence-Weighted Annotation Aggregation algorithm} for our annotations (see Algorithm~\ref{alg:conf_agg_condensed}). The algorithm employs a confidence-weighted inductive strategy (A1). Each annotation comprises a 3D spatio-temporal region, an artifact label, a confidence score and an optional textual reason. The aggregation process begins by sorting annotations in descending order of confidence. It then iteratively merges regions with significant overlap, quantified using 3D Intersection over Union (IoU)~\cite{ravi2020pytorch3d}. We empirically set the IoU threshold ($IoU\_{thresh}$) at 40\% to balance label granularity and semantic coherence. When merging, the \textbf{AppendAnnotationByLabel} function adds an annotation's confidence and reason to an existing aggregated region if the label matches. Otherwise, it assigns a distinct label (A2). We use confidence-weighted region aggregation (\textbf{ConfWeightedAvg}) to calculate the merged region's coordinates as a weighted average of contributing regions. Unlike Non-Maximum Suppression (NMS), our approach retains collective intelligence by weighting contributions rather than solely selecting the highest confidence region, thus mitigating risks from potential errors in top-confidence annotations. Annotations failing the $IoU_{thresh}$ criterion with all existing regions initiate new aggregation clusters (\textbf{CreateNewAggregatedRegion}). 

Following the merging phase, the algorithm calculates aggregated attributes. The final label of each aggregated region is determined as the label with the highest weighted confidence score ($r.\text{aggInfo}[\text{label}].\text{score}$). This score integrates the user's submitted confidence with their historical annotation confidence (retrieved via \textbf{GetUserHist}), averaged across participants (A2). This approach considers both self-assessed confidence and user capability derived from historical performance (A1, A2, G3). The intuition is that consistent high-confidence annotations might indicate higher accuracy. Other aggregated attributes include the aggregated confidence score, averaged across the top-T (empirically set T=5) highest individual confidence scores to ensure sufficient support from multiple high-confidence users for artifact identification. Finally, the algorithm aggregates textual reasons at the label level using language model-based embeddings (\textbf{LM\_Agg}), similar to prior work~\cite{yen2024memolet} (details in Appendix~\ref{sec:embedding}), to mitigate negative individual errors and promote positive social influence (A3). 

\begin{algorithm}
\caption{Confidence-weighted Aggregation} 
\label{alg:conf_agg_condensed} 
\begin{algorithmic}[1] 
\Require Annotations $A = \{a_i\}$, $IoU_{thresh}$, $T$ 

\State $A_{sorted} \gets \text{SortDesc}(A, \text{by confidence})$
\State $R \gets []$

\Comment{Phase 1: Merge annotations}
\ForAll{$a \in A_{sorted}$}
    \State $merged \gets \text{False}$
    \ForAll{$r \in R$}
        \If{$\text{IoU}(a.\text{region}, r.\text{region}) \ge IoU_{thresh}$}
            \State AppendAnnotationByLabel(r, a) 
            \State $r.\text{annotations}.\text{append}(a)$
            \State $r.\text{region} \gets \text{ConfWeightedAvg}(r.\text{annotations})$
            \State $merged \gets \text{True}$; \textbf{break}
        \EndIf
    \EndFor
    \If{not $merged$}
        \State $newAggR \gets \text{CreateNewAggregatedRegion}(a)$
        \State $R.\text{append}(newAggR)$
    \EndIf
\EndFor

\Comment{Phase 2: Calculate aggregated attributes}
\ForAll{$r \in R$}
    \ForAll{label, data in $r.\text{labelData}.\text{items()}$} \Comment{Iterate through each label in the region}
        \State $confs \gets [\text{pair.confidence for pair in data}]$
        \State $reasons \gets [\text{pair.reason for pair in data}]$ 
        \State $wScores \gets []$
        
        \ForAll{pair in data}
            \State $hist \gets \text{GetUserHist}(\text{pair.user})$ 
            \State $wScores.\text{append}(\text{pair.confidence} \times \text{Avg}(hist))$
        \EndFor
        
        \State $r.\text{aggInfo}[\text{label}].\text{score} \gets \text{Avg}(wScores)$ 
        \State $r.\text{aggInfo}[\text{label}].\text{conf} \gets \text{Avg}(\text{Top}(T, confs))$ 
        \State $r.\text{aggInfo}[\text{label}].\text{reason} \gets \text{LM\_Agg}(reasons)$ 
        
    \EndFor
\EndFor

\State \Return $R$
\end{algorithmic}
\end{algorithm}

\subsubsection{Label Demonstration Design}\label{sec:demonstration}

\begin{figure}[!htbp]
    \centering
    \includegraphics[width=0.5\textwidth]{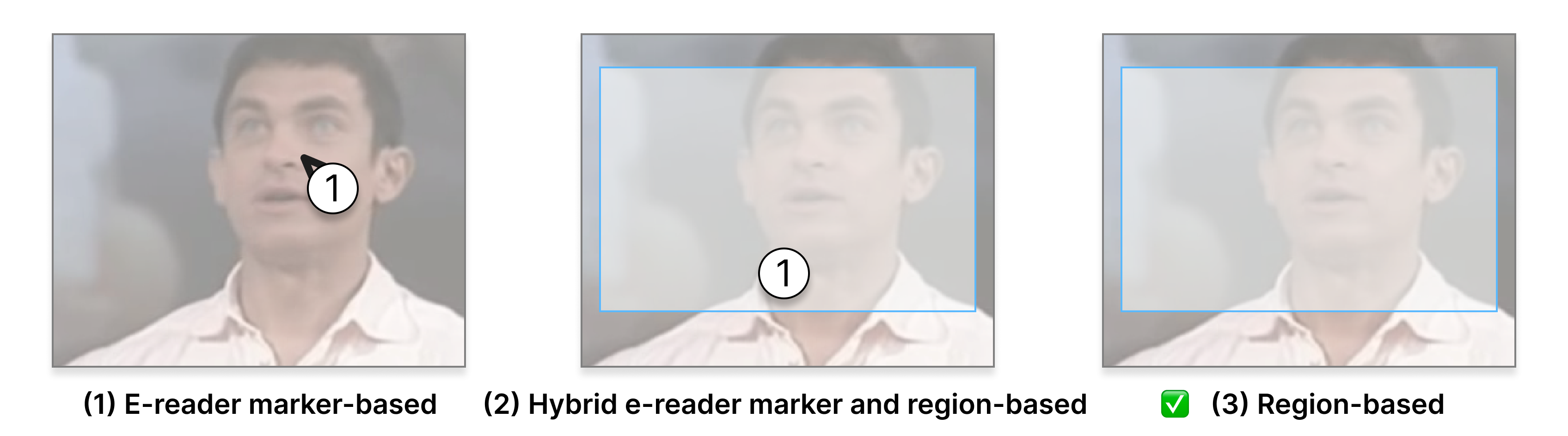}
    \caption{Candidate artifact demonstration designs. (a) E-reader Marker-based representation, (b) Hybrid e-reader marker/region-based representation, and (c) Region-based bounding box representation. Details are presented upon selection. The \textcircled{1} represents a marker.}
    \label{fig:demonstration}
\end{figure}

To mitigate negative social influence and foster critical evaluation (G2,G4,D1-D3), similar to the annotation process, we adopted a \textbf{bullet screen} metaphor (Figure~\ref{fig:demonstration} \textcircled{b}) for label demonstration. We selected the bullet screen metaphor over two alternative methods: an \textbf{e-reader style interface} (Figure~\ref{fig:demonstration} \textcircled{a}) and a \textbf{hybrid marker/region method} (Figure~\ref{fig:demonstration} \textcircled{c}). The dynamic overlay of bullet screen metaphor is well-suited for \textbf{foster social awareness and influence (G2)}. Furthermore, its visualization is integrated and less disruptive compared to alternatives, aligning with the considerations to \textbf{enhance user awareness (D1) without inducing cognitive overload (D2)}.

\textbf{Initial Cues and Interaction} Peer-contributed annotations are initially presented as subtle, semi-transparent highlights overlaying the relevant video regions. These highlights serve as unobtrusive cues~\cite{kirsh2000few}, inviting exploration of the aggregated peer annotations without disrupting the primary viewing experience or imposing undue cognitive load, aligning with the independent contribution principle (L2). The initial visualization deliberately conceals specific labels and confidence scores to discourage premature heuristic judgments by the user (L2). The demonstration of others' labels inherently motivates social contribution (L3). We refrained from restricting the number of displayed regions because deepfake identification often needs detecting multiple, distinct spatial artifacts. Preserving these diverse spatial indicators is therefore important for comprehensive coverage~\cite{joslin2024double}.

\textbf{Hierarchical Information Disclosure} Subsequently, hovering over an interactive highlighted region triggers a pop-up displaying candidate aggregated labels and the aggregated confidence score in descending order, based on the weighted confidence derived from the previous aggregation process. We empirically determine to list at most five labels to foreground the primary label effectively while mitigating cognitive overload~\cite{taky2024cognitive,robinson1994computational} (G2,D2). We also did not display exhaustive details in the initial pop-up view~\cite{kirsh2000few} to avoid cognitive overload. 

Further inquiry is supported by clicking on either a displayed label or its confidence score. To specifically mitigate potential conformity bias~\cite{kirsh2000few}, this action intentionally avoids revealing raw individual opinions. Instead, the interface presents \textit{thematic clusters of opinions} (Figure~\ref{fig:teaserfigure} \textcircled{4}), guiding user interpretation toward collective understanding. We segmented and clustered original opinions into aggregated concise textual summaries via language models to facilitate reflection~\cite{tanprasert2024debate}. This hierarchical disclosure strategy, which concurrently presents multiple labels accompanied by brief explanations, is designed to mitigate anchoring and framing biases~\cite{furnham2011literature,nelson1997toward} by preventing a disproportionate focus on initial or singularly prominent information.

\subsection{Implementation of Collab}

We implemented \proj{} as a web-based plugin using JavaScript for the front-end interface. The plugin is operating system independent, requiring only a modern web browser that supports JavaScript and HTML Canvas API (e.g., current versions of Chrome, Firefox, Edge, Safari). JavaScript handles user interactions, including annotation mode toggling, time segment definition, and label submission. It renders annotations via a transparent canvas layer overlaid on the video player. We verified the plugin's operational compatibility across several mainstream video-centric social media platforms, such as YouTube, $\mathbb{X}$ and Bilibili, confirming its applicability.

\textbf{Frontend and Backend Logic.} The system uses the Fetch API for asynchronous communication with the Python backend, where we immediately transmit annotation data upon submission. The Python backend, utilizing the Flask web framework, manages API requests and performed annotation aggregation. We use NumPy for numerical computations within the aggregation process, including confidence-weighted IoU calculations. 

\textbf{Demonstration Presentation.} To generate the visual demonstration overlays, the backend calculates aggregated confidence ($C_{agg}$) and inter-annotator agreement ($A_{agg}$) metrics. It then maps these to colors based on pre-defined thresholds: Green indicates high consensus ($C_{agg} \ge 75\%$ and $A_{agg} \ge 80\%$), while Red indicates low confidence or low agreement ($C_{agg} \le 40\%$ or $A_{agg} \le 50\%$), with Orange covering intermediate cases. The frontend renders these colors as semi-transparent overlays (40\% opacity). Full metric definitions and threshold details are in Appendix~\ref{sec:color}.

\textbf{Rationale Analysis.} For processing free-text rationales, the backend employs a pre-trained sentence transformer\footnote{\url{https://huggingface.co/sentence-transformers/all-MiniLM-L6-v2}} model to generate 384-dimensional semantic embeddings for each justification. These embeddings are then grouped using K-Means clustering (empirically set to K=5 clusters) to identify and present common underlying reasons within the interface. Appendix~\ref{sec:embedding} provides details on the model setup.

\section{Evaluating Collab in a Simulated Social Media Environment}

To address RQ3 and assess the effectiveness of \proj{} in identifying deepfake videos, we conducted a user study comparing \proj{} with alternative techniques on a simulated social media platform. Following prior literature~\cite{jahanbakhsh2022leveraging,jahanbakhsh2024browser}, we set the experiment duration at seven days. 

\subsection{Experiment Material and Platform}

The experiment material included videos from four prominent deepfake video datasets: Face Forensics++~\cite{rosslerfaceforensics}, BioDeepAV~\cite{croitoru2024deepfake}, DFW~\cite{pu2021deepfake} and DDL~\cite{miao2025ddl}. The selection was guided by three key criteria: (1) enabling a rigorous comparison of AI recognition versus human annotation accuracy, where these datasets are commonly benchmarked with AI algorithms~\cite{rosslerfaceforensics,croitoru2024deepfake,pu2021deepfake,miao2025ddl}, (2) broadening thematic diversity to cover different aspects of deepfake video analysis, and (3) prioritizing datasets with real-world content to enhance user engagement and relevance. To balance experimental scope with participant load considerations given our recruitment constraints, we randomly sampled 30 real and 30 fake videos from each dataset, totaling 240 videos, a number similar to prior practices~\cite{jahanbakhsh2022leveraging,jahanbakhsh2024browser}. This balanced distribution of real and fake videos facilitates the exploration of \proj{}'s algorithm and the corresponding user annotation behavior. Since our aggregation is activated by user annotations, its robustness is best evaluated by its capacity to filter false positives rather than omissions. Consequently, the 1:1 setting serves as a stress test regarding whether the \proj{} is robust to potential user hyper-vigilance and high-frequency erroneous inputs. Notably, this setting also aligns with prior practices in accuracy benchmarking~\cite{roitero2020can,roitero2023can,xu2022does}, although it diverges from the typical low prevalence in social media environments.

We mimicked the layout of $\mathbb{X}$ (formerly Twitter) and developed an online social media platform as the experiment platform, due to ethical concerns that preloading videos directly onto $\mathbb{X}$ may poison the ecological environment (see Figure~\ref{fig:interface}). Participants were introduced to the platform and allowed to familiarize themselves with it in advance. 

\subsection{Experiment Design and Procedure}\label{sec:design}

This study utilized a between-subjects design, with \textit{technique} as the between-subjects factor. We compared Collab with two alternative techniques, inspired by previous work~\cite{jahanbakhsh2022leveraging,jahanbakhsh2024browser} and specifically tailored to video misinformation annotation context (see Figure~\ref{fig:interface} for interface illustration): 

\begin{figure*}[!htbp]
    \centering 
    \subfloat[Collab.]{
        \includegraphics[width=0.32\textwidth]{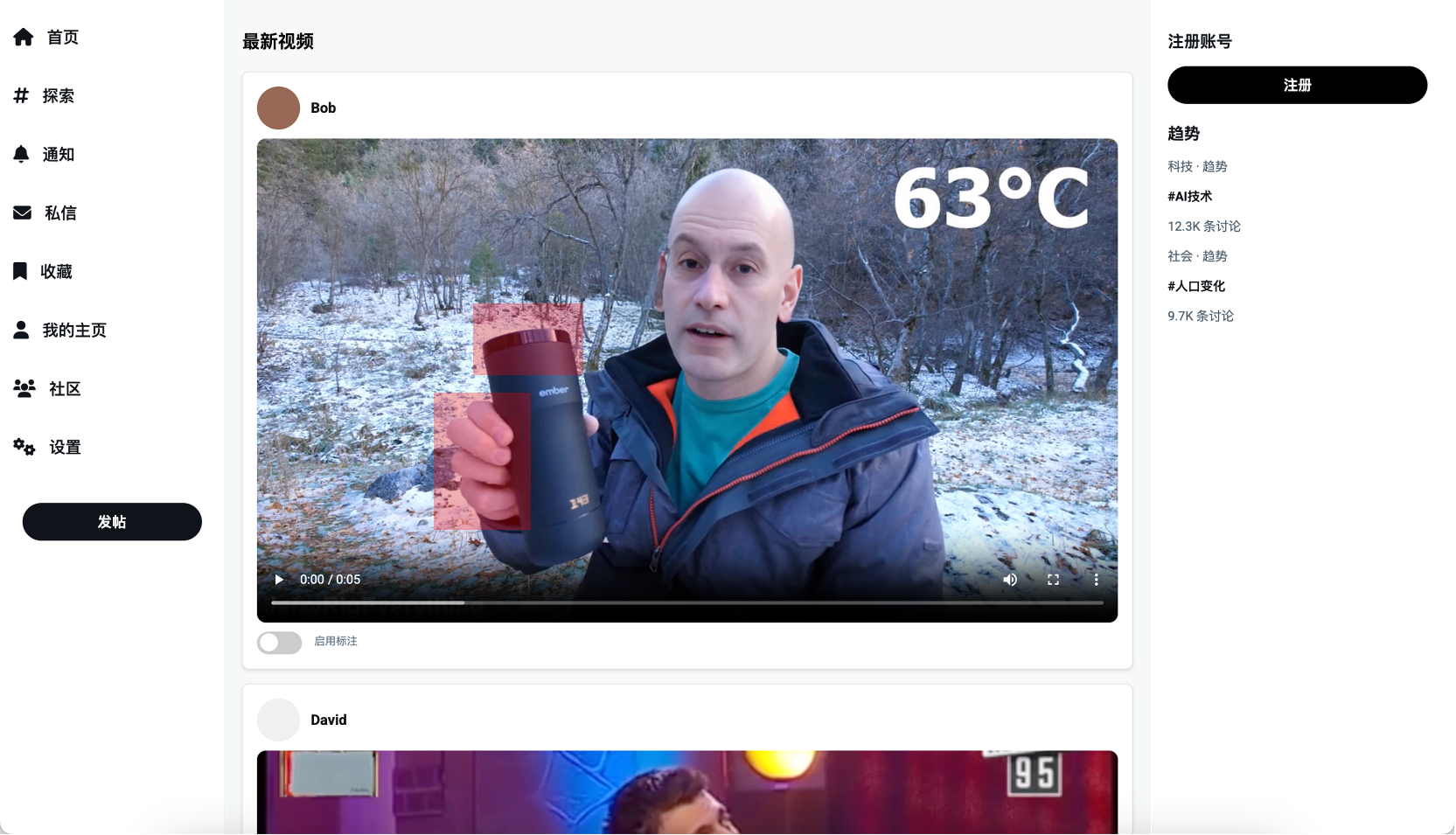}
    }
    \subfloat[No Agg.]{
        \includegraphics[width=0.32\textwidth]{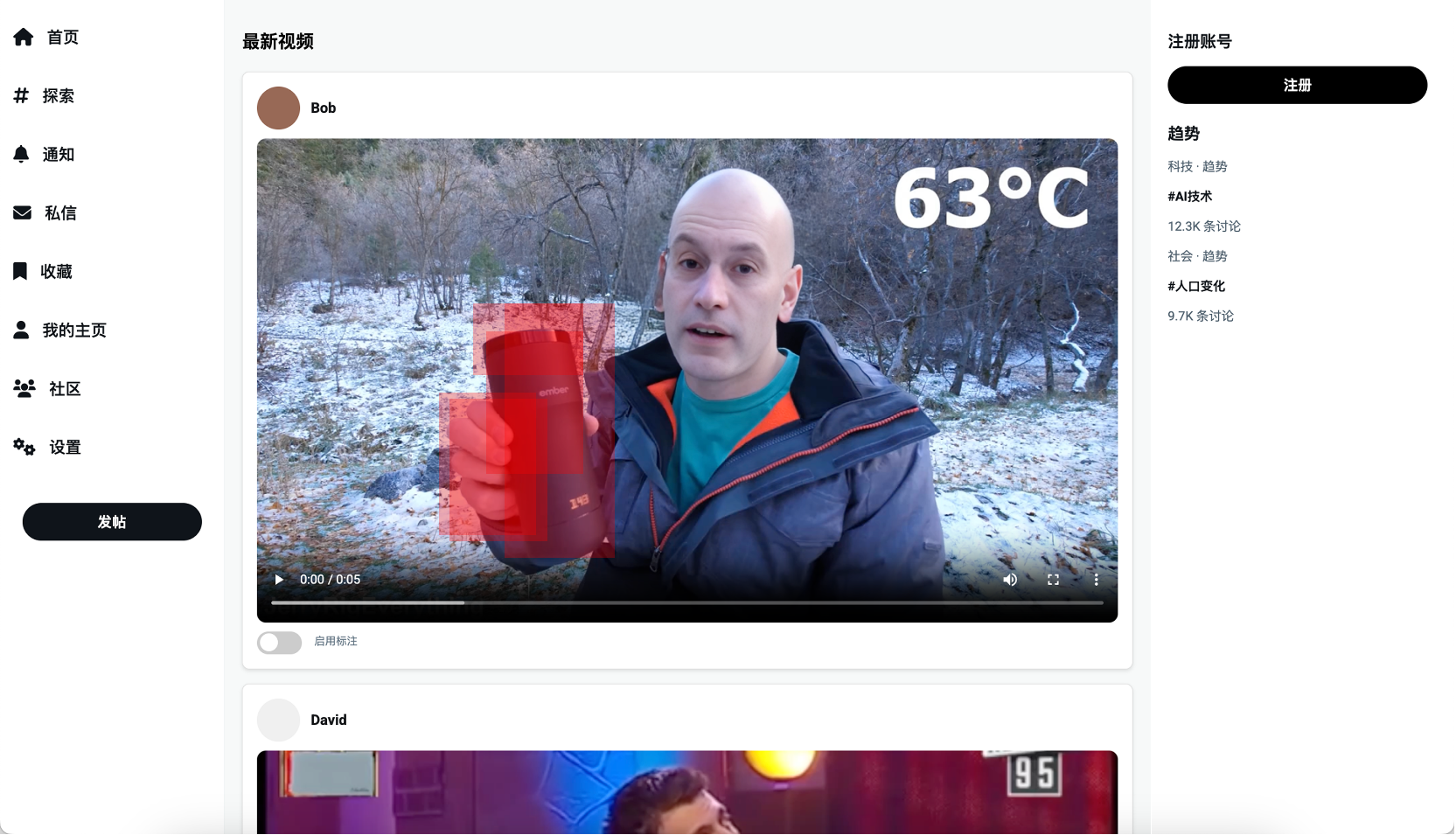}
    }
    \subfloat[No Label.]{
        \includegraphics[width=0.32\textwidth]{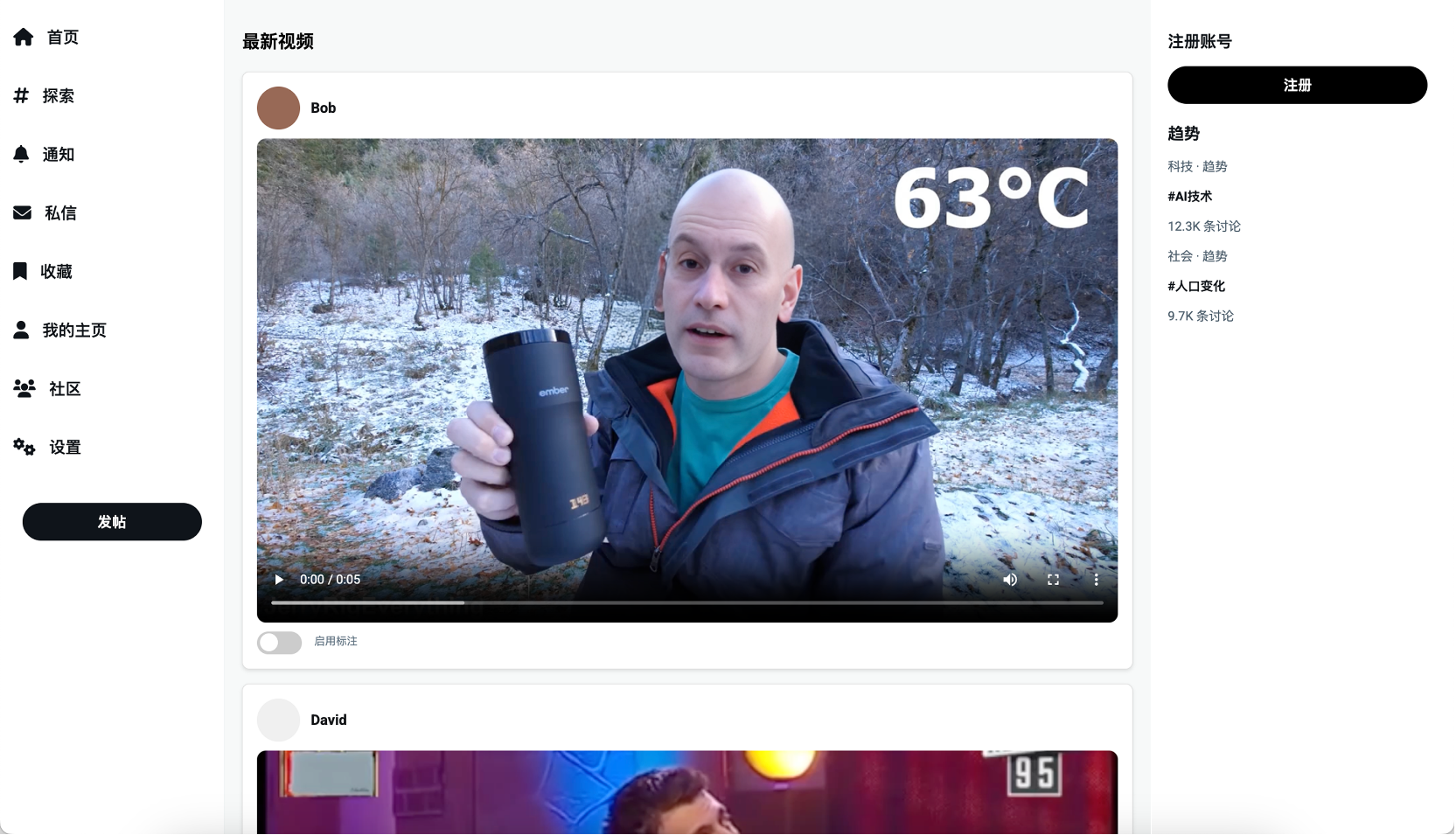}
    }
    \caption{The interface illustration for different techniques, where \textit{No Agg} shows non-aggregated labels, and \textit{No Label} shows no labels.}
    \label{fig:interface}
\end{figure*}

\begin{itemize}
    \item Collab: it is implemented as in Section~\ref{sec:design_implementation}.
    \item No Agg: The annotation process is similar to Collab, but users' annotations were not aggregated. Consequntly, the demonstration showed each user's raw annotation separately, similar to previous systems~\cite{jahanbakhsh2024browser}. However, the demonstration form was adapted to be video-based, mirroring \proj{}, because the original form was targeted at textual annotations and was not suitable for deepfake videos.
    \item No Label: The annotation and aggregation process is the same as Collab, but other users' annotations were not shown on the videos. 
\end{itemize}

Participants were evenly distributed across the three technologies. Each day, participants were instructed to browse the custom-built social media platform (see Figure~\ref{fig:interface}), and annotate any content that they deemed suspicious. While this explicit prompting diverges from naturalistic browsing, it was employed to ensure a basic level of interaction for evaluating \proj{}'s collaborative mechanics. Notably, participants were allowed to flag content they perceived as questionable, even in the absence of absolute certainty, to capture a spectrum of suspicion levels. We did not disclose the distribution of the videos, nor did we explicitly instruct them to annotate deepfake videos. Instead, we asked them to identify potentially fake, manipulated, or videos containing misinformation, to mimic a realistic setting. To model realistic usage patterns, all videos were preloaded into an infinite scroll interface of the platform similar to $\mathbb{X}$, allowing participants to browse the content through a natural, continuous feed. They could perform annotations in multiple sessions throughout the day. They needed to annotate at least five videos each day, determined according to previous literature~\cite{jahanbakhsh2022leveraging,jahanbakhsh2024browser} to balance users' fatigue and align with their daily viewing practices. Visual indicators were applied to completed videos in each participant's view to signal their annotation status. The order of the videos was randomly shuffled between participants, but remained consistent for each individual across the 7 days.

When they finished each daily session, they completed a survey to measure their subjective experience. This included the NASA-TLX questionnaire and 7-point Likert scales assessing \textit{reflectiveness}, \textit{rationality}, \textit{support}, \textit{overall satisfaction}, \textit{fluency}, \textit{novelty}, and \textit{effectiveness} (see Appendix~\ref{sec:questionnaire_for_study_2} for detailed definitions). Furthermore, we conducted semi-structured interviews on the first and seventh days to examine changes in participants' attitudes and strategies. The interviews explored their annotation behavior, their interaction with the technique, their experience with viewing and interpreting others' annotations, and suggestions for improvement. After the study, all participants were debriefed on the research goals and videos' distribution.

\subsection{Recruitment and Participants}

Participants were recruited via social media platforms and academic networks. Our recruitment involved 90 participants (51 males, 39 females, with a mean age of 23.3, SD=6.2), with a mix of expertise levels ranging from novice users to experts in media literacy, information science, and AI. Participants were each compensated 210RMB for their time, and informed consent was obtained before participation. The study was approved by our university's Institutional Review Board (IRB). No participants dropped out of the experiment. 

\subsection{Data Analysis}

We analyzed quantitative data according to its distribution. We employed a one-way Analysis of Variance (ANOVA) for metrics that conformed to a normal distribution, such as accuracy and confidence scores, followed by Tukey's HSD for post-hoc comparisons. For the ordinal data from subjective ratings on 7-point Likert scales, which did not meet normality assumptions, we used the non-parametric Kruskal-Wallis test, with Dunn's test and a Bonferroni correction for post-hoc analyses. For qualitative data, we conducted a thematic analysis~\cite{clarke2017thematic}. One primary author coded 20\% of the responses to develop an initial codebook. This author then discussed with a secondary author, and refined the codebook to resolve disagreements before applying it to the entire dataset.

\subsection{Results}\label{sec:results}

Our evaluation showed that \proj{}'s effectively enhanced video misinformation identification accuracy and fostered user engagement, aligning with our core design goals. Over a 7-day period, participants annotated an average 110.4 videos in 7-day's duration (SD=46.7), with \proj{} receiving the highest annotation number (M=190.8, SD=49.2) and No-label received the the lowest (M=62.1, SD=10.6). This suggested that participants were intrinsically motivated to annotate more videos than required. Furthermore, no participant reported fatigue during the interviews, suggesting the task was not perceived as burdensome. For participant identification in the results section, we used identifiers (e.g., G1P1) to denote the condition (G1: Collab, G2: No Agg, G3: No Label) and participant's number. 

\subsubsection{Annotation Accuracy}

We defined accuracy based on the factual correctness of participant annotations against the ground truth label in the dataset. As \proj{} captures a spectrum of confidence rather than binary labels, establishing a classification threshold could distinguish tentative suspicion from definitive detection. We selected a confidence threshold of 80\% combining qualitative and quantitative evidence. Qualitatively, participants consistently indicated in the interview that a confidence level of 80\% of higher represented definite judgments, while lower scores reflected uncertainty. Quantitatively, we conducted a sensitivity analysis (Appendix~\ref{app:sensitivity_analysis}), confirming that the 80\% threshold yielded the highest F1-score by lowering false positives associated with lower-confidence annotations. Therefore, an annotation was categorized as ``fake'' if the associated confidence rating was 80\% or higher. Conversely, annotations with a confidence level below 80\%, or videos that were not annotated at all, were classified as ``real''. 

With this threshold, \textit{\proj{}} significantly increased the factual correctness of video identification, achieving an F1-score of 0.883 (Table~\ref{tbl:collab}). This represents a substantial improvement over the \textit{No Label} condition ($\Delta = 0.088$ for F1-score) and a modest gain over the \textit{No Agg} condition ($\Delta=0.034$ for F1-score). Statistical testing confirmed a significant difference between techniques ($F_{2,87} = 3.454$, $p = .036 < .05$), with post-hoc analysis revealing \proj{}'s superiority over \textit{No Label} ($p < .05$). This outcome supports the efficacy of the aggregation (A1) and demonstration (D1,D2,D3) strategies in synthesizing individual inputs into reliable collective judgments.

\begin{table}
\centering
\caption{The precision (P), recall (R) and F1-score (F) of Collab and other alternative techniques. Notably, the dataset is balanced, with the number of fake videos equaling that of the authentic ones. Avg. denoted the averaged accuracy across the four datasets.}
\label{tbl:collab}
\resizebox{0.5\textwidth}{!}{%
\begin{tabular}{lccccc}
\toprule
P/R/F & Face Forensics++ & BioDeepAV & DFW & DDL & Avg. \\
\midrule
No Label & \cellcolor{lightBarBlue!60}0.692/0.967/0.806 & \cellcolor{lightBarBlue!30}0.700/0.712/0.706 & \cellcolor{lightBarBlue!100}\textcolor{white}{1.000/0.915/0.956} & \cellcolor{lightBarBlue!35}0.750/0.677/0.712 & \cellcolor{lightBarBlue!55}0.786/0.818/0.795 \\
No Agg & \cellcolor{mediumBarBlue!60}0.787/0.903/0.841 & \cellcolor{mediumBarBlue!45}0.826/0.769/0.797 & \cellcolor{mediumBarBlue!100}\textcolor{white}{1.000/0.970/0.985} & \cellcolor{mediumBarBlue!40}0.769/0.784/0.776 & \cellcolor{mediumBarBlue!65}0.845/0.856/0.849 \\
Collab & \cellcolor{darkBarBlue!75}0.838/0.900/0.868 & \cellcolor{darkBarBlue!75}0.829/0.900/0.863 & \cellcolor{darkBarBlue!100}\textcolor{white}{1.000/0.983/0.992} & \cellcolor{darkBarBlue!60}0.816/0.800/0.808 & \cellcolor{darkBarBlue!80}0.871/0.896/0.883 \\
\bottomrule

\end{tabular}
}%
\end{table}

\begin{figure}  
    \centering
    \includegraphics[width=0.5\textwidth]{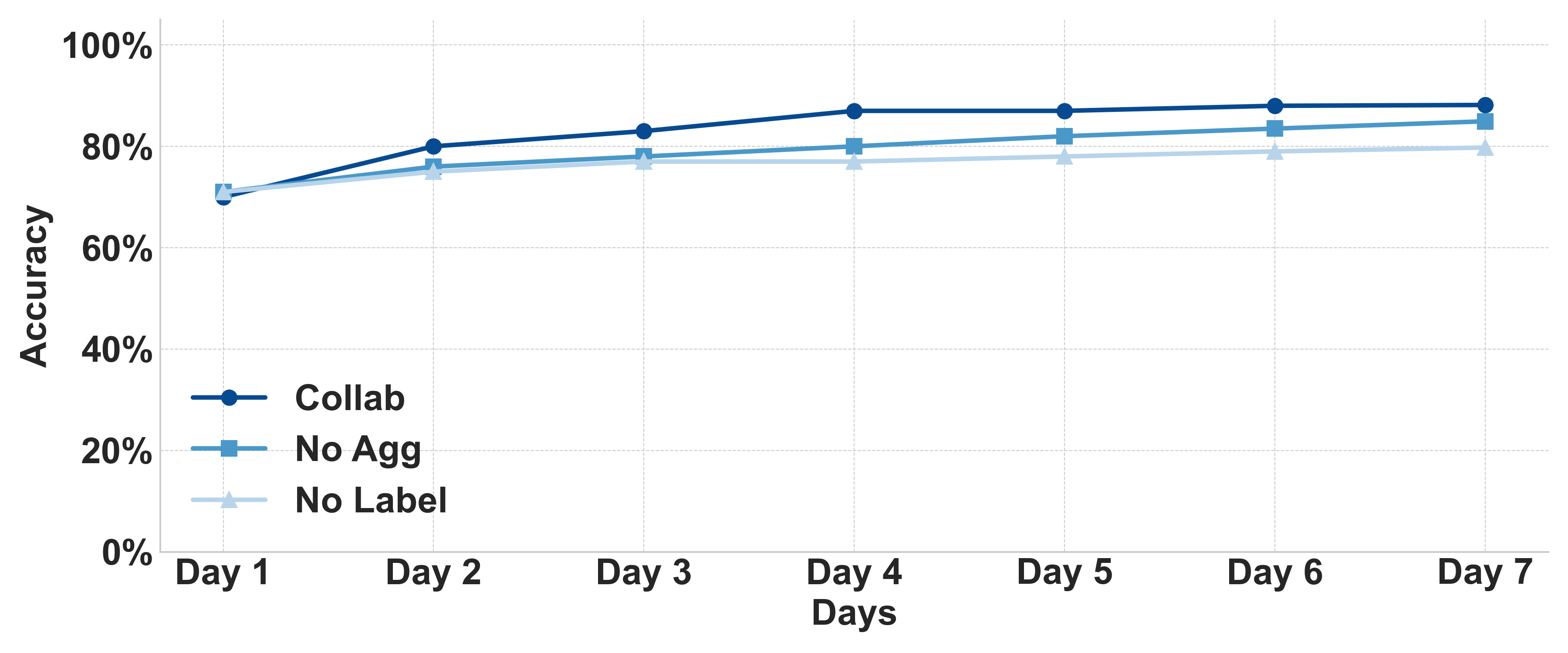}
    \caption{Average accuracy of different techniques across seven days.}
    \label{fig:technique}
\end{figure}

Accuracy improved over the 7-day study period for all techniques, with \proj{} showing rapid gains initially and reaching saturation around Day 4 (Figure~\ref{fig:technique}). In comparison, \textit{No Agg} technique's accuracy reached 84.10\% even on Day 7, increasing daily while with a low speed. \textit{No Label} technique reached 79.00\% accuracy in Day 7, although it converged faster, as others' labels did not influence the users' annotations. Statistical testing revealed significant effect of time on accuracy for different techniques (\proj{}: $F_{2,87} = 3.341$, $p < .05$; \textit{No Agg}: $F_{2,87} = 3.727$, $p < .05$; \textit{No Label}: $F_{2,45} = 3.145$, $p < .05$), although we did not find post-hoc differences between different days across techniques. 

\subsubsection{Annotation Distribution}\label{sec:distribution}

Following established reporting practices~\cite{xu2024dipa2}, the analysis of annotation characteristics, confidence ratings, bounding box geometry, and label distributions revealed that \textit{\proj{}} fosters users to critically view videos and input fine-grained annotations.

\begin{figure}
    \centering
    \includegraphics[width=0.5\textwidth]{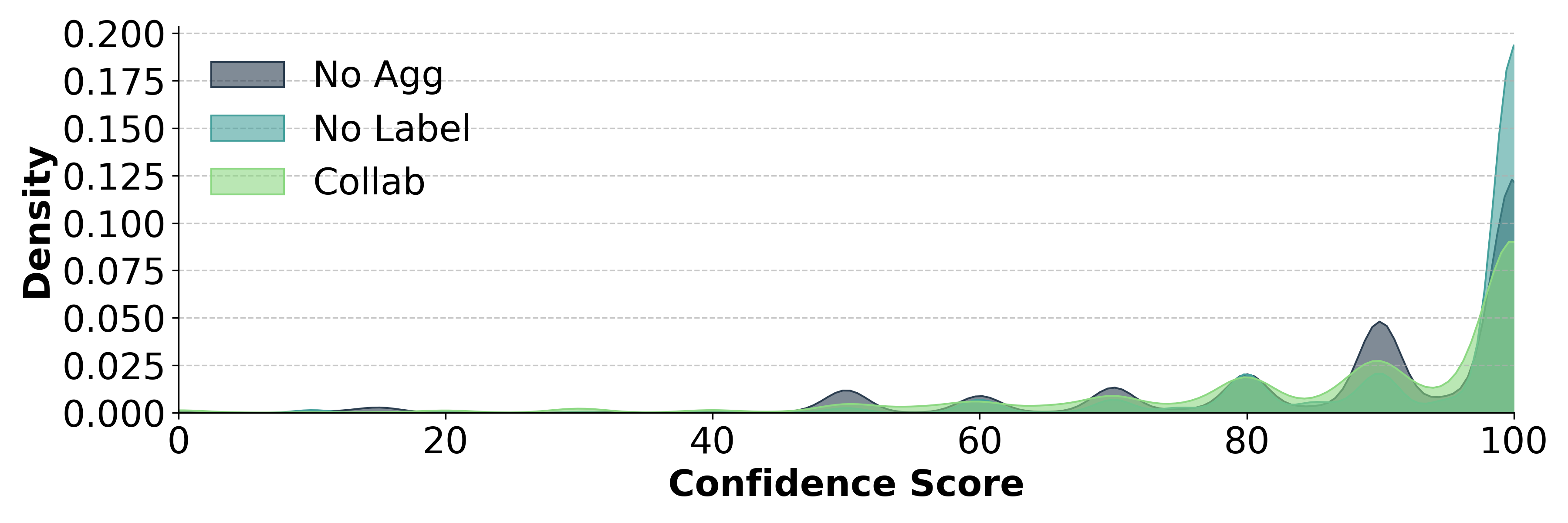}
    \caption{The Kernel Density Estimation (KDE) plot illustrating the confidence distribution of the annotations for different techniques.}
    \label{fig:confidence}
\end{figure}

\textbf{Annotation Confidence:} User confidence ratings provide insight into perceived certainty and the system's capacity to support accurate judgment. As depicted in Figure~\ref{fig:confidence}, distinct confidence patterns emerged across the experimental conditions. Overall, over 90\% of the confidence scores are above 60\%, indicating that participants generally felt confident about their labeling, a finding corroborated by interview results that most participants thought they could discern that a video was fake. We found that \textit{\proj{}} has the lowest mean confidence at 86.9\% (SD=13.7\%), while No-label has the highest mean confidence at 93.44\% (SD=18.04\%). The \textit{No Agg} condition has a mean confidence at 88.1\% (SD=19.3\%). Contrary to intuition, we found users in \textit{No Agg} usually would feel confident about their labeling even when their labels may not be correct. Therefore, the lowest mean confidence in \proj{} suggests that it may encourage users to think cautiously. After careful deliberation, they were able to assign a more rational and accurate confidence score, rather than a deceptively high score that could mislead others. We also found significant effect of technique on the ratings ($F_{2, 87} = 99.460$, $p < .001$). 

\textbf{Bounding boxes:} We found that technique significantly influenced users' annotation behavior, affecting both the size and spatial distribution of annotations. We found a significant effect of the technique on the area of the bounding boxes ($H = 220.457$, $p < .001$). Annotations made with \textit{\proj{}} were substantially smaller and more precise (median area = 1.94\%, M = 5.48\%, SD = 9.11\%) compared to those in the \textit{No Agg} (median = 3.27\%, M = 7.51\%, SD = 11.05\%) and \textit{No Label} conditions (median = 5.87\%, M = 11.29\%, SD=13.22\%). Post-hoc tests confirmed that \textit{\proj{}} led to significantly smaller annotations than both baseline conditions ($p < .001$). This suggests that the collaborative signals in \textit{\proj{}} enabled users to move beyond identifying large, general areas of suspicion and instead focus on finer-grained, more specific visual artifacts. 

Furthermore, technique also significantly affected the spatial positioning of annotations. We observed a significant effect of technique on the average horizontal position ($H = 16.926$, $p < .001$) and vertical position ($H = 99.571$, $p < .001$). Annotations in the \textit{\proj{}} condition were, on average, located differently (M = 36.9\% on X-axis, M=28.3\% on Y-axis) than those in the No-agg (M=40.2\% on X, M=31.0\% on Y) and No-label (M=42.5\% on X, M=35.8\% on Y) conditions. These results indicate that \textit{\proj{}} not only changed the granularity of users' annotations but also guided their visual attention to different regions of the video. Taken together, these findings suggest that \textit{\proj{}} altered users' analytical process, fostering a precise and focused approach to identifying potential manipulations.

\begin{figure*}  
    \subfloat[No Agg.]{
        \includegraphics[width=0.32\textwidth]{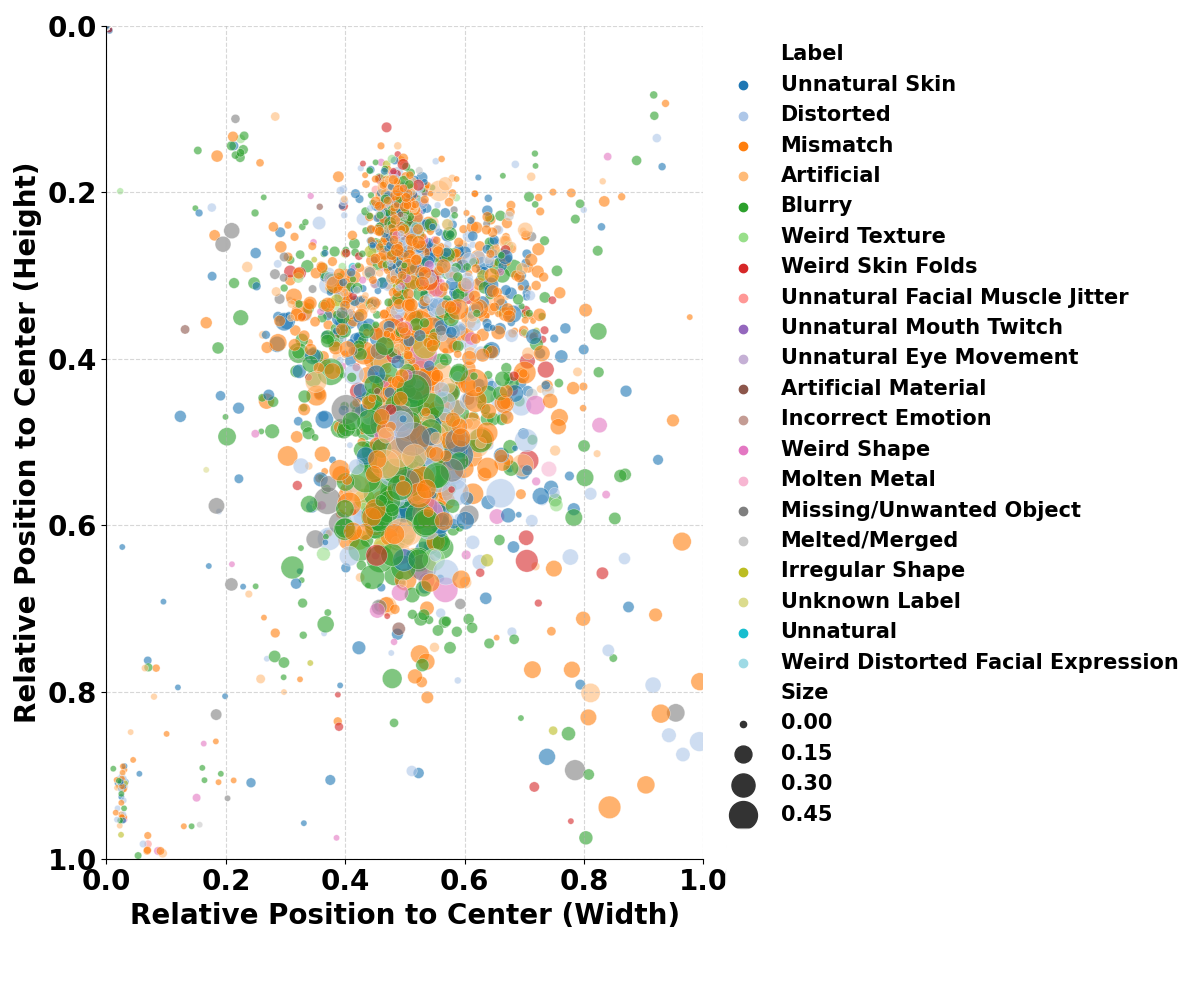}
        \label{fig:label-dist-no-agg}
    }
    \subfloat[No Label.]{
        \includegraphics[width=0.32\textwidth]{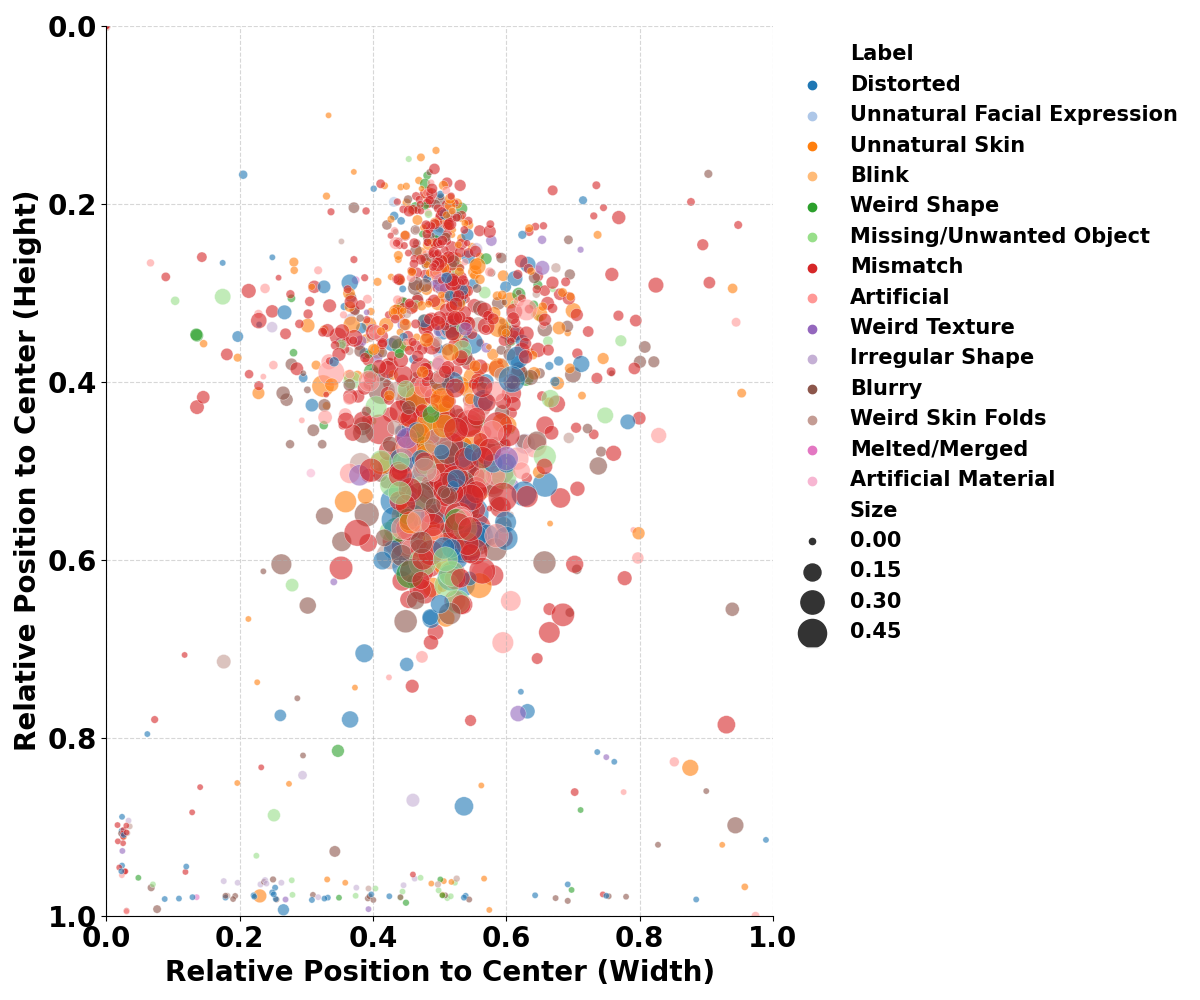}
        \label{fig:label-dist-no-label}
    }
    \subfloat[Collab.]{
        \includegraphics[width=0.32\textwidth]{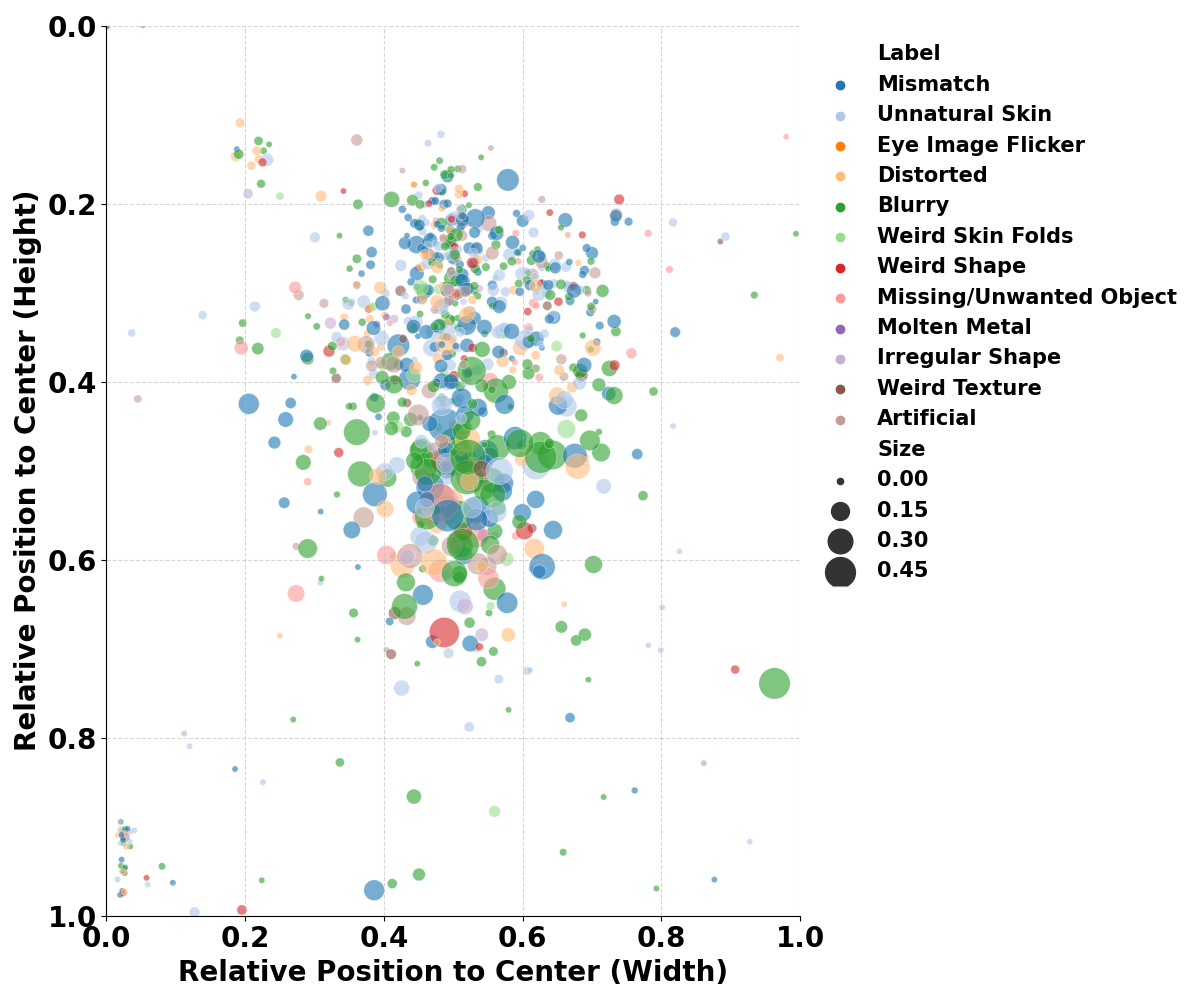}
        \label{fig:label-dist-collab}
    }
    \caption{The distribution of labels for different techniques. The size of the point indicated the size of the annotation.}
    \label{fig:label}
\end{figure*}  

\textbf{Label distribution:} We found a strong preference for the predefined categories over creating custom ones. Interviews corroborates this finding, with participants consistently reporting that the provided labels were sufficient to describe the artifacts they identified, rather than an inability to formulate new descriptions. This is reflected in the data, where custom labels constituted less than 0.5\% of all annotations across all techniques. The few instances of custom labels included \textit{unnatural facial muscle jitter} (0.26\%) in the \textit{No Agg} condition and \textit{unnatural facial expression} (0.22\%) in the \textit{No Label} condition. The use of custom labels was lowest in the \textit{\proj{}} condition, with only \textit{eye image flicker} (0.06\%) being used.

For the distribution of the predefined labels, in the alternative conditions, \textit{Mismatch} was the most frequent label (44.80\% and 36.55\%, respectively). This suggests that without specific guidance, users often rely on \textit{Mismatch} as a general-purpose category to articulate a sense of incongruity, even if they cannot pinpoint the precise visual artifact causing it.

This pattern shifted significantly in the \textit{\proj{}} condition. The usage of the general \textit{Mismatch} label dropped substantially (to 23.51\%), while annotations for concrete, evidence-based artifacts saw a marked increase. Specifically, \textit{Blurry} became the most prominent label (28.26\%), and the use of \textit{Unnatural Skin} also grew (to 20.52\%). This indicates that the collaborative signals in \textit{\proj{}} may have effectively reduced user uncertainty. By highlighting peer labelings, \textit{\proj{}} empowered participants to think beyond a general feeling, and instead identify and label specific evidence.

\textbf{Temporal evolution and conformity analysis:} To address concerns regarding potential bias, we further analyzed the temporal evolution of annotation behaviors (detailed in Appendix~\ref{app:temporal}). Results indicated that while the spatial overlap of annotations increased over the 7-day period ($p < .001$, calculating 3D IoU per annotation compared with earlier aggregated boxes), the average bounding box size significantly decreased ($p < .001$), and label vocabulary shifted toward high specificity. These opposing trends--increasing consensus accompanied by increasing granularity--suggest that \textit{\proj{}} facilitates conformity with thoughtful reasoning rather than blind conformity. Users appear to utilize peer annotations as attentional cues for detailed inspection, resulting in a refined, rational consensus.

\subsubsection{Subjective Ratings}\label{sec:subjective_rating}

\begin{figure*}  
    \centering 
    \subfloat[NASA-TLX.]{
        \includegraphics[width=0.7\textwidth]{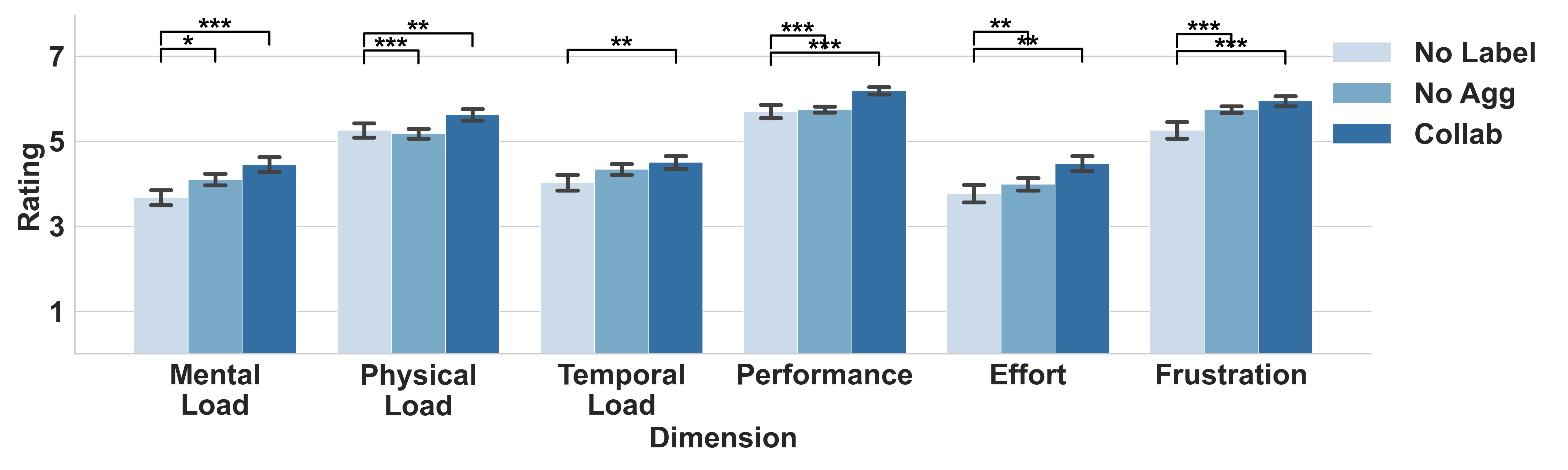} 
        \label{fig:nasa_tlx}
    }

    \subfloat[Other subjective rating dimensions.]{
        \includegraphics[width=0.7\textwidth]{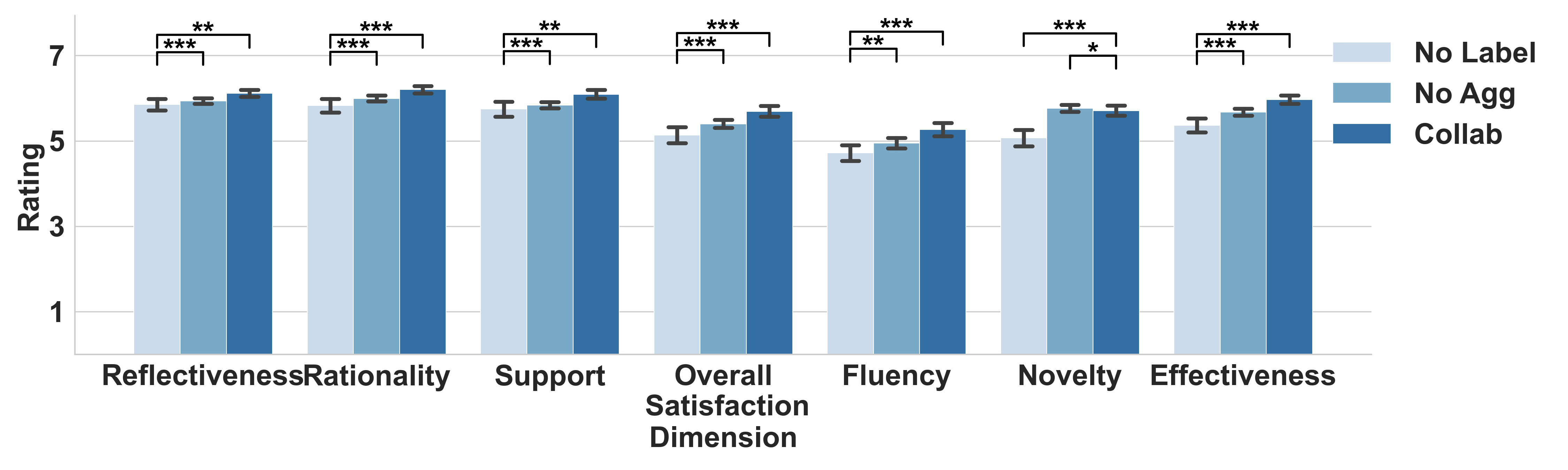}
        \label{fig:other}
    }
    \caption{The subjective ratings across different techniques (1: most negative, 7: most positive). Errorbar indicated one standard deviation. *, **, *** indicated significance at $p < .05$, $p < .01$, $p < .001$ levels separately.} 
    \label{fig:subj} 
\end{figure*}

Our analysis of subjective ratings reveals a clear and consistent preference for \proj{} across a wide range of metrics, including workload (i.e., NASA-TLX) and user experience, as illustrated in Figure~\ref{fig:subj}.

An analysis of NASA-TLX dimensions (Figure~\ref{fig:nasa_tlx}) showed that technique had a significant effect on all six dimensions. Post-hoc comparisons consistently showed that \textit{\proj{}} significantly reduced the perceived task demands compared to baseline conditions. For instance, participants rated \textit{\proj{}} as requiring significantly less mental load ($H = 17.162$, $p < .001$, $\eta^2 = .050$) and physical load ($H = 20.763$, $p < .001$, $\eta^2 = .061$) than both \textit{No Agg} and \textit{No Label}. These findings suggest that the collaborative signals provided by \textit{\proj{}} effectively offloaded cognitive and physical effort. Similarly, \textit{\proj{}} was perceived as requiring less effort ($H = 14.375$, $p < .001$, $\eta^2 = .042$) and inducing substantially less frustration ($H = 25.453$, $p < .001$, $\eta^2 = .075$). Participants also reported a significantly higher level of performance when using \textit{\proj{}} ($H = 40.673$, $p < .001$, $\eta^2 = .120$). For temporal load, \textit{\proj{}} was also rated more favorably than \textit{No Label} ($H = 9.125$, $p = .010$, $\eta^2 = .027$), indicating an efficient user experience. 

This strong preference for \textit{\proj{}} extended to broad user experience dimensions (Figure~\ref{fig:other}). We found a significant effect of technique on all subjective dimensions. The results indicated that \textit{\proj{}} was not only felt easier to use but also better supported the users' analytical process. It was rated as significantly more effective at encouraging reflectiveness on video authenticity ($H = 22.166$, $p < .001$, $\eta^2 = .065$) and promoting rationality in judgment ($H = 23.497$, $p < .001$, $\eta^2 = .069$) compared to both baselines. Furthermore, participants found \textit{\proj{}} provided superior support for their task ($H = 23.929$, $p < .001$, $\eta^2 = .070$). Unsurprisingly, these benefits culminated in significantly higher ratings for overall satisfaction ($H = 23.067$, $p < .001$, $\eta^2 = .068$), system fluency ($H = 17.131$, $p < .001$, $\eta^2 = .050$), and perceived effectiveness ($H = 34.425$, $p < .001$, $\eta^2 = .101$). On the dimension of novelty, while \textit{\proj{}} was seen as more novel than \textit{No Agg} ($p < .001$), we also observed that \textit{No Agg} was rated as significantly more novel than \textit{No Label} ($p = .031 < .05$), suggesting that any form of analytical support was perceived as a novel intervention by participants. 

\subsubsection{Qualitative User Perceptions}\label{sec:qualitative_feedback}

We collected and reported participants' feedback about their annotation strategy, process, their experience, opinions on the collaborative annotation and potential improvements. 

\textbf{Annotation locus and detection heuristics.} Participants demonstrated a strong consensus in their annotation focus, primarily concentrating on the human face as the primary site of manipulation. As G1P8 stated, \textit{``These past two days, I've generally been marking the faces of the people in the videos.''} Within the facial region, annotations were localized to specific features where anomalies were most salient, such as the eyes, mouth, and jaw. G2P10 noted, \textit{``The first is the eyebrows. The second is the eyes''}, and G2P19 similarly remarked, \textit{``It's mostly the facial features, for example, the eyes, the normal face, lips, and jaw''}.

This focus was guided by a consistent set of detection heuristics. The most prevalent was the identification of splicing artifacts and unnatural textures, where a face appeared artificially overlaid, described by one participant as looking like it was \textit{``pasted on''} (G3P7). Another key heuristic was sensitivity to unnatural motion and expressions, such as \textit{``facial twitching''} (G3P4) or abnormal blinking patterns. G1P8 observed that in some videos, \textit{``the blinking frequency is abnormal. Sometimes they don't blink at all throughout a video, or sometimes the blinking frequency is too high.''} Annotations were also triggered by logical inconsistencies that violated real-world expectations, such as gender-mismatched features. G1P4 considered this a \textit{``very obvious sign of AI,''} citing examples where characters had \textit{``a man's body and a woman's appearance.''} A final, less common heuristic involved lighting and shadow anomalies, with G3P5 mentioning \textit{``unnatural transitions in light and shadow''} as a reason for annotation. While the predefined labels like \textit{``unnatural skin''}, \textit{``distortion''}, and \textit{``mismatch''} were frequently used, some participants found them too generic. G1P30 felt that \textit{`` `mismatch' is a bit too broad''}, and G1P8 expressed a desire for more specific custom labels like \textit{``splicing issues''} to better capture the observed flaws.

\textbf{Annotation strategies and workflow.} To manage the cognitive demands of the task, participants developed distinct annotation strategies and workflows. A dominant approach was a two-pass method, where users would first watch a video to gain a holistic impression before re-watching it to place precise annotations. G3P7 articulated this strategy: \textit{``I will watch it once first, and after watching it, I will find the problematic parts on the second pass and mark them.''} Over time, however, some users transitioned to a more fluid, real-time workflow, opting to \textit{``catch the glitch as it happens.''} (G3P29) This method also helps to mitigate negative conformity bias, where the re-watching process allowed users to reflect on their own judgment criteria and compare to others' judgment critically.

A clear divergence in behavior emerged regarding the confidence threshold for placing an annotation. A conservative group would only mark anomalies they were highly certain about. As G3P27 stated, \textit{``I tend to mark things that I'm 100\% certain about or that are obviously problematic at a glance.''} In contrast, other participants were willing to document their suspicions with less certainty, using the system's confidence slider to reflect their doubt. G3P7, for example, would mark less obvious issues with a confidence of \textit{``50 or 60''}. However, all participants consistently used high scores ($\ge$ 80\%) to denote definite judgments, while using lower scores to flag potential but uncertain anomalies. The scope of the annotation also varied, with some participants drawing large bounding boxes around the entire face, while others adopted a more precise approach. G2P28 highlighted this difference in strategy: \textit{``I usually focus on the details. For instance, everyone might circle the entire facial features, but I might just circle a lip or an eye.''}

\textbf{The impact and management of social annotations.} For participants exposed to peer contributions, these social signals may act as a double-edged sword. For example, G2P19 noted a potential drawback was \textit{``not thinking for oneself and just using others' [ideas].''} Furthermore, the visual overlay of bounding boxes was sometimes cited as a hindrance, especially frequent in non-aggregated group, that was \textit{``very disruptive to the viewing experience''} (G2P11), because it could obscure the very details a user was trying to scrutinize. However, empirical observation and feedback from participants in the \proj{} group suggests that the visual load remained manageable, as the number of concurrent bounding boxes per frame peaked at only three. We therefore interpret the obstruction by participants as a productive friction that disrupts passive consumption and compels active scrutiny. Some users in \textit{No Label} groups also worried that seeing others' marks \textit{``might influence my own judgment,''} (G3P4), a concern that they were already vigilant and prepared. Consequently, the majority of participants utilized social annotation constructively via four distinct mechanisms:

\textbf{\textit{Acting as attentional cues:}} Social annotations served to direct user attention to regions they might have otherwise overlooked. Participants reported that seeing another's mark could \textit{``remind me''} of subtle flaws they had initially missed.

\textbf{\textit{Confirming independent hypotheses:}} Rather than replacing user judgment, peer labels often acted as validators. Users noted that overlapping labels helped to \textit{``confirm that my own idea is correct,''} increasing their confidence in their assessment.

\textbf{\textit{Fostering critical examination (resolving conformity bias):}} Counteracting conformity bias, users employed a verification strategy where they performed independent analysis before consulting social information. As G1P8 explained, \textit{``I will look after I have finished my own annotation''}. This process prompted critical reflection, with users intentionally using others' labels to \textit{``think about why they were marking this way''} rather than passively accepting them. These strategies facilitate the consensus derived from independent validation rather the blind compliance.

\textbf{\textit{Novelty seeking and utility addition (gamification effect):}} High consensus on existing labels motivated some users in the late period of the study to engage in novelty seeking. Rather than redundantly agreeing with existing highlights, these users sought to add marginal utility to the system by finding \textit{``new details that others may had missed''}. G1P9 described this motivation: \textit{``As others have already found these labels... I prefer to highlight new artifacts un-explored''} These behavior facilitates the expansion of collective detection coverage, avoiding premature convergence to dominant features.

\textbf{System usability and potential improvements.} Across the interviews, participants provided consistent feedback on the system's usability and expressed clear preferences for future features. The most frequently cited usability issue was the inability to control the video's progress bar while in annotation mode, which complicated the process of locating a specific frame. G3P4 expressed this frustration, wishing to \textit{``let the annotator drag the progress bar themselves.''}

A strong desire emerged for more user control over the display of social annotations. A common suggestion was to implement a toggle switch, allowing users to show or hide others' marks. G2P11 proposed, \textit{``you could add an option for them to check a box to turn on others' annotations or turn them off.''} Another popular suggestion was to move annotation details to a non-occluding side panel. A final major point of friction was the system's lack of state persistence. Users found it difficult to resume their work after a page refresh or on a subsequent day, as the system did not save their progress. G2P12 described this problem: \textit{``today my computer restarted... when I re-entered, I couldn't find my previous marks... I didn't know where I had seen up to before.''} This lack of browsing history was identified as a key area for improvement.

\section{Discussions}

\subsection{Feasibility}

\textbf{Accuracy.} When using \proj{}, participants achieved an accuracy of 88.2\% in identifying deepfake videos. This performance significantly surpassed alternative conditions that lacked the system's core aggregation or demonstration mechanisms (see Table~\ref{tbl:collab}). This accuracy is particularly compelling when contextualized against the performance of algorithmic detection methods benchmarked on the same datasets. For instance, the \proj{}'s accuracy on the DFW dataset (F1-score 0.992) is higher than the 0.81 F1-score of the winning DFDC model~\cite{dolhansky2020deepfake} and the 0.91 F1-score of the Capsule-DA method~\cite{pu2021deepfake}. Similarly, its performance on the FaceForensics++ dataset (F1-score 0.868) is comparable with the 87.81\% accuracy reported by XceptionNet on compressed videos~\cite{rosslerfaceforensics}. While different metrics preclude a direct comparison on BioDeepAV dataset, \proj{}'s F1-score (0.863) contrasts with the leading reported AUCs of 0.5852 for MRDF~\cite{zou2024cross} and 0.6195 for StA~\cite{yan2025generalizing}. Collectively, these comparisons suggest that \proj{} may harness collective human intelligence to achieve performance feasible of content moderation.

\textbf{Motivation and scalability.} A critical consideration for \proj{} is users' motivation, as labeling requires investment of time and cognitive effort~\cite{jahanbakhsh2022leveraging,jahanbakhsh2024browser}. To address this challenge, \proj{} reframes this demanding task by integrating it into the familiar and engaging paradigm of bullet-screen commenting. This design probably explains why, in contrast to prior work~\cite{jahanbakhsh2024browser}, our participants perceived the task as fast to complete. The high volume of annotations generated (see Section~\ref{sec:results}) serves as evidence off this approach's success in fostering user engagement.

Our findings indicate that users may not perceive this activity as ``uncompensated labor''~\cite{jahanbakhsh2022leveraging}, but rather as a form of civic participation or self-expression. Aligning with prior work ~\cite{jahanbakhsh2022leveraging,jahanbakhsh2024browser}, this intrinsic motivation appears to stem from a sense of accomplishment, community engagement, and a desire to protect their social circles. To further foster engagement, \proj{} could incorporate extrinsic motivators in the future. Formally recognizing high-quality contributions, similar to the certification systems on $\mathbb{X}$, or implementing reputation points could leverage social contribution (L3)~\cite{foskett1955social} to incentivize sustained participation. 

Regarding scalability, \proj{}'s reliance on voluntary contributions means it can be strategically deployed. A pragmatic approach, similar to that of $\mathbb{X}$'s Community Notes, would be to prioritize high-traffic videos that are more likely to become viral and cause widespread harm~\cite{hughes2024viblio}. To address the long tail of low-traffic content, \proj{} could be complemented by automated detection or human-AI collaborative systems~\cite{jahanbakhsh2023exploring,zhang2025exploring}, presenting a promising model for overcoming the ineffectiveness of many existing interventions~\cite{hartwig2024adolescents,hartwig2024landscape}. However, such hybrid approaches must be implemented carefully, as they may introduce challenges related to user trust in automated agents~\cite{zhang2024profiling}.

\textbf{Critical Analysis.} Our findings in Section~\ref{sec:qualitative_feedback} suggest that \proj{} shifts users from passive consumption to active critical analysis, addressing passivity of users in combating misinformation~\cite{gamage2022designing,violot2024shorts}. To support this, \proj{} employs a scalability strategy distinct from text-based platforms like $\mathbb{X}$'s Community Notes~\cite{chuai2024community}. While Community Notes restricts display to a single textual explanation to minimize clutter, deepfake detection often includes multiple subtle spatio-temporal artifacts~\cite{joslin2024double}. Therefore, rather than filtering to a single output, our algorithm (Section~\ref{sec:aggregation}) retains multiple distinct clusters and consolidated raw inputs into visual consensus regions. This preserves the granularity necessary for users to spot manipulation artifacts, such as a \textit{localized flicker} or an \textit{unnatural skin texture}. Meanwhile, this strategy avoids overwhelming users, as participant feedback in Section~\ref{sec:qualitative_feedback} suggests that the bounding boxes do not lead to overwhelming information overload or excessive occlusion.

Regarding visual overload in real-world deployments, we suggest adopting filtering mechanisms similar to $\mathbb{X}$'s Community Notes. While our study shows that users tend to converge on finite, salient artifacts (e.g., facial glitches), future iterations could restrict visibility to the top-k regions based on consensus or confidence scores. However, we emphasize that such filtering algorithms must be robust against adversarial manipulation, including coordinated attacks designed to skew annotations~\cite{truong2025community}.

Finally, we interpret mixed user feedback regarding social annotations (Section~\ref{sec:qualitative_feedback}) as indicative of productive cognitive friction~\cite{cox2016design} rather than a design flaw. In contrast to the \textit{No Agg} condition, which users in Section~\ref{sec:qualitative_feedback} criticized for obtrusive occlusion, \proj{} streamlines the label identification process yet retains sufficient friction to provoke critical analysis. This friction successfully disrupted the ``truth-default'' mindset~\cite{epstein2023social} inherent in passive scrolling. Indeed, despite a few cites around visual load, participants voluntarily contributed more annotations than required (see Section~\ref{sec:results}). Ultimately, this engagement proved empowering rather than burdensome, as users perceived high utility in the interactive verification process (Section~\ref{sec:subjective_rating}).


\textbf{Cold-start.} A potential challenge for \proj{} is the ``cold-start'' problem, where new videos lack the initial annotations required to enable the system's core aggregation and demonstration features. However, our experimental findings suggest that this issue may be less severe in practice, as \proj{} demonstrated the ability to adapt rapidly with only a few seed labels (see Figure~\ref{fig:collab_design}). To further mitigate this challenge, we suggest implementing several bootstrapping strategies. For instance, initial ``seed'' annotations could be provided by expert fact-checkers or a group of trusted users with high reputation scores. Alternatively, the process could be initiated using automated or human-AI collaborative systems~\cite{jahanbakhsh2023exploring}. Such systems could generate a preliminary annotation layer for users to confirm or refine, ranging from simply detecting salient regions like faces as a baseline~\cite{joslin2024double} to complex analyses. It is crucial to note, however, that these automated approaches risk introducing initial biases that could influence subsequent user contributions~\cite{jahanbakhsh2023exploring}.

\textbf{Ecological Validity.} Our methodology described in Section~\ref{sec:design} differs from real-world contexts in two key aspects. \textbf{First, regarding the balanced dataset,} while real-world misinformation is sparse, we intentionally balanced the distribution to reliably benchmark recall and detection boundaries~\cite{roitero2020can,roitero2023can,xu2022does}. We acknowledge that organic scarcity often reinforces a ``truth-default'' mindset~\cite{epstein2023social}, or conversely leads to over-sensitivity towards fake content~\cite{mink2024s}, potentially leading to reduced recall and higher precision. In contrast, our setting likely induced heightened vigilance. However, this design mirrors specific high-exposure environments, such as ``Echo Chamber''~\cite{cinelli2021echo}, where users are surrounded with high ratio of fake videos. With our setting, future work could also consider deployment scenarios where the system acts as a second-stage filter following algorithmic pre-filtering. Such pre-filtering could created a roughly balanced real-to-fake ratio for human reviewers similar to our experimental conditions, though we recommend that future work should also explore fully naturalistic settings~\cite{jahanbakhsh2022leveraging,jahanbakhsh2024browser}.

\textbf{Second, we employed explicit annotation prompts to mitigate data sparsity.} In short-term studies, relying on spontaneous engagement risks data sparsity, which may prevent the accumulation of social signals needed to trigger the collaborative loop. By ensuring baseline interactions, we evaluated \proj{}'s efficacy.  Crucially, we operationalized these prompts using neutral language that encouraged scrutinizing `suspicious' content and annotating, instead of mandating the detection of manipulation. Consequently, the observation that participants voluntarily exceeded the mandatory minimum annotation number in Section~\ref{sec:results} indicates genuine intrinsic motivation and perceived utility rather than mere compliance. This suggests our setup effectively simulates a scenario similar to community fact-check~\cite{chuai2024community}, implying that real-world deployment may be effective among moderators or fact-check volunteers to overcome the inertia of passive consumption. While we prioritized such phrasing to ensure validity, we recommend that future work systematically examine how varying prompt formulations might further influence annotation behavior and accuracy~\cite{konstantinou2025behavior}.

\subsection{Generalizability}

We discuss the generalizability of \proj{} across the broad scope of misinformation. Beyond binary deepfake detection, \proj{} facilitates deepfake localization and interpretation~\cite{miao2025ddl}, where \proj{}'s fine-grained labels could provide nuanced explanations. This spatio-temporal paradigm extends to cheapfakes, especially those manipulated through photoshopping and speed changes, where annotations could target  manipulated visual artifacts and temporal changes. Regarding out-of-context videos, while \proj{} allows users to anchor skepticism by annotating visual discrepancies, comprehensive debunking in this domain may require integration with fact-checking workflows to supply external context~\cite{hughes2024viblio}. Furthermore, the framework holds potential for multimodal verification, such as flagging voice cloning or lip-sync errors~\cite{wang2025fighting}. This extension is contingent upon interface adaptations to interactive audio waveform annotations alongside visual tracks.

Beyond the video domain, the collaborative workflow principles of \proj{} offer potential for adaptation to other media formats. Core mechanisms, such as preserving the independence of initial annotations and using confidence-weighted aggregation, are not inherently media-specific. These principles could be applied to analyze static images or textual content that contains manipulated information, such as doctored photographs or misleading infographics~\cite{castano2019leveraging,suzuki2015poster}. However, such adaptation is non-trivial and necessitates significant interface modifications. For instance, it requires replacing temporal controls with spatial annotation tools for images or sentence-level markers for text.

Finally, \proj{} could be conceptualized not as a standalone tool but as a component within broader, hybrid fact-checking workflows. User-generated annotations, while scalable, can be susceptible to bias. This limitation motivates an integration where \proj{} functions as a first-pass system to triage content and direct the attention of professional fact-checkers to high-risk items. In such models, crowd-generated labels and expert assessments could be displayed concurrently, leveraging the scalability of the crowd while grounding the final judgment in expert analysis~\cite{martel2023misinformation}. Exploring this fusion of crowd intelligence with expert oversight offers a promising direction for developing resilient and scalable systems that draw upon the complementary strengths of both human-centric and human-AI collaborative frameworks \cite{jahanbakhsh2022leveraging,jahanbakhsh2021exploring}.


\subsection{Potential Harm}

Like other crowdsourcing platforms~\cite{hughes2024viblio,jahanbakhsh2024browser}, \proj{} must be equipped against risks of poor annotation quality, systemic bias, and adversarial manipulation.

A primary concern is the integrity of annotations and the system's vulnerability to malicious actors. Our study suggests that \proj{}'s design has some inherent resilience. The confidence-weighted aggregation algorithm and users' tendency toward critical evaluation can marginalize isolated, inaccurate annotations, protecting collective judgment against manipulation attempts~\cite{le2020malcom,wilder2019defending}. However, this defense may fail against coordinated disinformation campaigns or sustained low-quality contributions that could discredit authentic content or fuel conspiracies. To address this, future iterations should integrate sophisticated governance mechanisms. Drawing from successful platforms like $\mathbb{X}$'s Community Notes and Wikipedia, we can implement reputation systems that weigh contributions based on users' historical accuracy and reliability~\cite{jahanbakhsh2024browser,communitynotes2022diversity}, establish clear contribution standards~\cite{wiki:2023citingsources}, use voting mechanisms to validate individual annotations~\cite{chuai2024community}, and deploy robust evaluation algorithms such as those in BirdWatch~\cite{prollochs2022community} or HawkEye~\cite{mujumdar2021hawkeye}. Adapting such mechanisms is particularly crucial, especially since \proj{}'s fine-grained spatial-temporal data is more complex to fact-check than static images or text~\cite{joslin2024double}.

A second challenge involves mitigating cognitive and social biases that can lead to misclassifications and mistrust. User overconfidence or implicit biases could result in the disproportionate flagging of content featuring marginalized communities, causing significant economic~\cite{myers2018censored} and emotional harms, such as feelings of oppression or invisibility~\cite{van2020shadowbanning}. Research indicates that moderators are less likely to misclassify users with whom they share an identity, consistent with in-group favoritism theories~\cite{mink2024s,pettigrew2006meta}. Therefore, mitigation strategies should include maintaining a diverse annotator pool, and potentially implementing consensus rules that require cross-demographic agreement, such as those on $\mathbb{X}$'s Community Notes system. Such approaches could also help repair user trust and counter the general mistrust arising from prolonged exposure to manipulated media~\cite{mink2024s,jahanbakhsh2024browser,jahanbakhsh2022leveraging}.

Furthermore, \proj{}'s governance must balance transparency with security. Full transparency of the aggregation algorithm could allow adversarial actors to reverse-engineer and ``game'' the system~\cite{bhuiyan2021nudgecred}. A potential solution is to introduce a degree of algorithmic ``fuzzing'', similar to Reddit's strategy for its voting system, which makes precise manipulation difficult~\cite{Coldewey2016}. However, this opacity carries risks, as it could be misused by platform developers to drive engagement or other corporate goals without accountability~\cite{Hao2021}. Consequently, any implementation of strategic opacity should be paired with strong external oversight and clear, enforceable policies to maintain system integrity and user trust.

\subsection{Design Implications}

Our findings inform the design of interactive systems that supported structured, collaborative video analysis. We propose three key design implications targeted distinct stages of the analysis workflow to enhance the overall reliability and efficiency: (1) the \textbf{\textit{interface design}} for individual structured data entry, (2) the \textbf{\textit{algorithmic design}} for aggregating collaborative data, and (3) the \textbf{\textit{interaction paradigms}} for managing information flow and cognitive bias.

\textbf{Implication 1. Integrate Structured Analytical Annotation Directly within Video Playback Interfaces.} To facilitate rigorous analysis, videos analysis tools should directly integrate mechanisms for structured annotation, including semantic labels, confidence scores, and textual rationales, onto the video timeline. This approach benefits multiple domains. For AI development, it enables the collection of rich, contextualized supervisory data, incorporating human judgment nuances like confidence and reasoning, crucial for training and evaluating sophisticated video understanding~\cite{madan2024foundation} and generation~\cite{sun2024diffusion} models. For collaborative scientific observation, it empowers research teams~\cite{ferligoj2015scientific} to precisely link interpretations and methodological notes to specific observed events, fostering shared understanding.

\textbf{Implication 2. Aggregate collaborative spatio-temporal annotations via confidence-weighted fusion.} To derive reliable insights from the collaborative analysis of ambiguous video content~\cite{ashton1985aggregating}, aggregation algorithms should fuse individual spatio-temporal annotations using confidence-weighted schemes. By systematically prioritizing contributions with higher reported confidence and employing spatio-temporally aware overlap metrics, this method produces a dependable group opinions' representation. Such an approach is particularly valuable in fields requiring the interpretation of complex dynamics, such as fine-grained behavioral coding~\cite{heyman2014behavioral} or nuanced human interaction analysis~\cite{jordan1995interaction}.

\textbf{Implication 3. Balance guidance and bias in collaborative video analysis via hierarchical disclosure.} Designers of interactive collaborative video analysis systems~\cite{yang2003interactive} should present aggregated peer annotations using a hierarchical disclosure strategy. This approach strategically navigates the tension between providing helpful guidance and introducing cognitive bias. It leverages collective insights to direct a user's attention toward relevant phenomena or areas of disagreement, while simultaneously mitigating the risk of premature consensus of conformity effects. This is especially critical for ambiguous data susceptible to interpretation variance, especially for domains like recommendation systems~\cite{wang2021user} and questionnaire design~\cite{gorrell2011countering}.
 
\section{Limitation and Future Work}

We acknowledge limitations in this work that outline directions for future research.

First, regarding ecological validity, our evaluation was conducted in a simulated environment with explicit annotation prompts, an equal number of real and fake videos, empirically derived accuracy thresholds, the absence of any aggregated annotation display limit and a specific participant demographic. While this controlled setting allowed us to focus on exploring \proj{}'s mechanisms and evaluating its accuracy, the current findings do not fully capture the complexities of real-world social media ecosystems, particularly regarding technical scalability and resilience against adversarial attacks. Future field studies are therefore needed to evaluate the system's robustness and accuracy. 


Second, the scope of interaction and evaluation presents opportunities for expansion. (1) Our study prioritized active annotators. Although we observed that these annotators showed improved reflection and examination even on non-annotated videos, future work should examine how passive users, who solely consume pre-labeled content, interpret these social signals. (2) \proj{} restricted inputs to flagging false videos. Future iterations could enable users to verify authentic content or commenting annotations to diversify input semantics. (3) Regarding mechanism isolation, future work should empirically compare \proj{} against alternative visualization strategies. This is crucial to disentangle the observed rational consensus from confounding factors such as social conformity, gamification, and the artifact's intrinsic perceptibility.

\section{Conclusion}

This paper introduced \proj{}, an annotation tool specifically designed for deepfake video identification via user collaboration. \proj{} seamlessly integrates labeling, aggregation, and demonstration processes, strategically leveraging principles from collective intelligence and social influence theories. Our design effectively translates core goals, including facilitating temporal and contextual interpretation, mitigating cognitive biases, ensuring interaction efficiency, and promoting reflective practices, into tangible features. These features encompass intuitive spatio-temporal annotation tools, a confidence-weighted 3D IoU algorithm for robust aggregation and a structured, non-intrusive visualization of collective insights. A 7-day online user study (N=90) evaluated \proj{} against alternative techniques without aggregation or demonstration separately. \proj{} demonstrated superior effectiveness in identifying misinformation, with 0.883 F1-score. The study also confirmed that \proj{} fosters users' reflective thinking and rational evaluation.



\begin{acks}
    This work is supported by the Natural Science Foundation of China (NSFC) under Grant No. 62132010. This work is also supported by Zhongguancun Laboratory.
\end{acks}

\bibliographystyle{ACM-Reference-Format}
\bibliography{sample-base}

\appendix 

\section{Ethical Implications}

We acknowledged that our paper has ethical concerns and tried our best to address these concerns. We followed Menlo report~\cite{bailey2012menlo} and Belmont report~\cite{beauchamp2008belmont} in organizing, designing and carrying out studies. Our study protocol received full approval from our institution's Institutional Review Board (IRB). To protect participants, all participants were informed of their right to withdraw from the study at any time without penalty or the need to provide a reason. They received appropriate compensation in accordance with local wage standards. At the designing stage, our study encouraged participants to discern deepfake videos and critically think about the deepfake artifacts. Participants also agreed unanimously they reflected more critically after using the systems. After the study, we debriefed the study's content, the fake videos to the participants and encouraged them to critically examine the deepfake videos. 

\section{Detailed Comparisons of Different Annotation Design Candidates}\label{sec:annotation_comparison}

From Figure~\ref{fig:collab_design}, we finally selected the \textcircled{3} and the by default annotation design strategy during region selection. This choice was informed by principles of CI and SI, and prioritize independent user judgment. 

\begin{itemize}
    \item For \textcircled{1}, the new user's labeling were set to other users' dominant labeling results by default as a nudge to guide users' selection.
    \item For \textcircled{2}, other users' dominant label were set as an anchor upon a new user's labeling and selection. 
    \item For \textcircled{3}, other users' labeling were refrained from showing, ensuring independent contribution and minimized negative social influence.
\end{itemize}  

We adopted \textcircled{3}. This decision is driven by the need for foster relatively independent critical assessment (G4) and minimize cognitive biases in collaborative settings (G2). Exposing users to dominant peer opinions before they form their own judgment (as in \textcircled{1} and \textcircled{2} significantly risks introducing anchoring effects~\cite{furnham2011literature} and conformity pressure~\cite{friedkin1998structural}. In deepfake video identification where nuances and diverse interpretations are vital, such biases could prematurely narrow the scope of evaluation, and suppress valuable, less common perspectives, compromising the quality of CI. Similarly, regarding the annotation design during region selection:

\begin{itemize}
    \item For \textcircled{4}, the other users' labeled regions could be shown upon users' annotation, with varying transparency based on distance, allowing user to explicitly adopt or align with them.
    \item For \textcircled{6}, the other users' labeled regions would be visualized as anchors with explicit distance cues, enabling users to ``snap'' their bounding boxes to existing ones. 
    \item By default, the other users' labeled regions were demonstrated separately, independently from users' annotations.
\end{itemize}

We selected the independent region visualization strategy (the by default strategy). This approach maintains a clear cognitive and visual separation between the task of defining a region based on their own perception and the task of viewing peer contributions. This separation supports independent initial judgment (L2, G2) and minimizes the cognitive load (G3) associated with interpreting complex visual cues (\textcircled{4} and \textcircled{6}).

\section{The Parameters of the Demonstration Phases Regarding Colors}\label{sec:color}

This section details the specific parameters used for the color-coded visualization of aggregated user annotations. The visualization renders aggregated annotation data as an overlay (e.g., heatmap or markers) on the video interface, designed to convey collective confidence levels and inter-annotator agreement for specific regions.

\subsection{Aggregation Metric}

For a given region or segment annotated by multiple users, two key metrics were computed from the collected annotations to drive the visualization:

\noindent \textbf{Aggregated Confidence $(C_{agg})$}: Represents the central tendency of confidence scores submitted by users for that specific region. In our implementation, it was calculated as the mean confidence score of all annotations overlapping the region.

\noindent \textbf{Inter-Annotator Agreement $(A_{agg})$}: Quantifies the level of consensus among annotators regarding the presence of type of deepfake videos in the region. This was calculated using the percentage of annotators assigning the most frequent label type to the region out of several candidates like the proportion of annotators who marked the region compared to those who viewed it.

\subsection{Color-Mapping Thresholds}

Based on the pilot study results, the following thresholds for $C_{agg}$ (ranging 0-100\%) and $A_{agg}$ (ranging 0-100\%) were established to map aggregated annotation data to distinct colors:

$\bullet$ \textbf{Green (High Confidence / High Agreement)}: Applied to regions where both the aggregated confidence and inter-annotator agreement were high, indicating strong consensus and certainty, with thresholds: $C_{agg} \ge 75\%$ and $A_{agg} \ge 80\%$. 

$\bullet$ \textbf{Red (Low Confidence / Low Agreement / High Uncertainty)}: Applied to regions characterized by low confidence annotations or significant disagreement / uncertainty among annotators, with thresholds: $C_{agg} \le 40\%$ or $A_{agg} \le 50\%$. 

$\bullet$ \textbf{Orange (Intermediate / Moderate Disagreement)}: Applied to all regions not meeting the criteria for Green or Red. This typically represented areas with moderate confidence levels, or areas where there was reasonable agreement but lower certainty, with thresholds: applied implicitly to regions where ($C_{agg} \ge 40\%$ and $A_{agg} \ge 50\%$) and not ($C_{agg} \ge 75\%$ and $A_{agg} \ge 80\%$). 

\subsection{Visual Representation}

The corresponding colors were rendered as semi-transparent overlays on the video. The opacity level was set to 40\% to ensure the underlying video content remained visible while clearly conveying the aggregated annotation information. The specific color hues used were standard Green (e.g., ``\verb|#00FF00|''), Orange (e.g., ``\verb|#FFA500|''), and Red (e.g., ``\verb|#FF0000|'').

\section{The Settings of the Embeddings}\label{sec:embedding}

We elaborate on the methodology employed for processing and categorizing the free-text justifications provided by users alongside their annotations. The primary objective was to group semantically similar justifications, enabling users to efficiently understand the common rationales behind aggregated annotations without cognitive overload.

\subsection{Language Model Selection and Embedding Generation}

We utilized a pre-trained sentence transformer model, specifically `sentence-transformers/all-MiniLM-L6-v2'\footnote{\url{https://huggingface.co/sentence-transformers/all-MiniLM-L6-v2}}. The model was selected for its balance of computational effciency and effectiveness in generating semantically rich embeddings for short textual inputs like user justifications. Each user-provided justification was fed into the selected model. The model processed the text and outputted a 384 dimensional fixed-size dense vector embedding representing the semantic meaning of the justification.

The generated vector embeddings served as the input for an automated categorization process. Embeddings were grouped using K-Means unsupervised clustering, with scikit-learn Python library. We used Cosine Similarity as the distance measure between embeddings within the clustering algorithm as it is well-suited for high-dimensional semantic vectors. The number of distinct categories was determined empirically as 5 because human could process limited number of on  blocks at one time and 5 could be appropriate amount~\cite{marois2005capacity}.

\section{Questionnaire Details for Study 2}~\label{sec:questionnaire_for_study_2}

We measured the following aspects using 7-point Likert scales:

\begin{itemize}
    \item \textbf{Motivation for Reflection (Reflectiveness)}: The extent to which the system prompted users to consider video authenticity.
    \item \textbf{Support for Rational Evaluation (Rationality)}: The degree to which the system facilitated reasoned assessment of video authenticity.
    \item \textbf{Support for Video Authenticity Assessment (Support)}: The system's perceived utility in aiding authenticity judgments during media consumption.
    \item \textbf{Overall Satisfaction (Overall satisfaction)}: Overall contentment with the system.
    \item \textbf{Perceived Fluency (Fluency)}: Ease of operation and interaction smoothness.
    \item \textbf{Perceived Novelty (Novelty)}: The system's perceived originality or uniqueness.
    \item \textbf{Perceived Effectiveness (Effectiveness)}: The overall belief in the system's ability to achieve its intended purpose.
\end{itemize}

\section{Sensitivity Analysis of the Confidence Threshold}\label{app:sensitivity_analysis}

A methodological challenge in evaluating \proj{} is mapping participants' nuanced, confidence-based annotations to the binary ``fake'' or ``real'' classifications required for accuracy metrics (e.g., precision, recall). Simply equaling any annotation as fake mapping would be imprecise, as our interviews revealed that users' mental models vary with confidence: low-confidence scores represented suspicion, whereas high-confidence scores represented a definitive judgment of fakeness. We therefore established a confidence threshold as our mapping mechanism.

\begin{figure}[!htbp]
    \centering
    \includegraphics[width=0.5\textwidth]{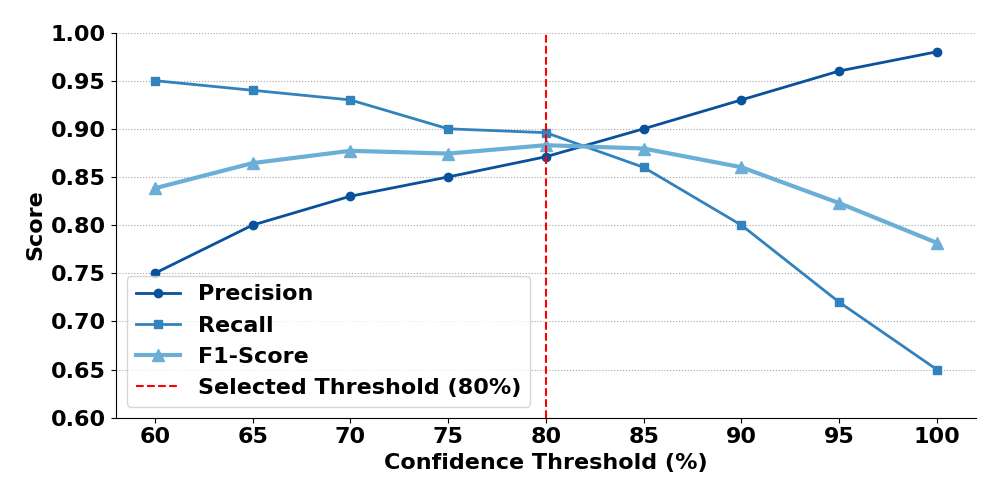}
    \caption{The sensitivity analysis, with the confidence threshold ranging from 60\% to 100\%. The red line denoted the situation with the confidence threshold as 80\%.}
    \label{fig:sensitivity}
\end{figure}

Therefore we conducted a sensitivity analysis. We calculated precision, recall and F1-score combining all datasets for the \proj{} condition, testing thresholds from 60\% to 100\% at 5\% intervals. As shown in Figure~\ref{fig:sensitivity}, this analysis revealed the expected precision-recall trade-off, where precision increased with the threshold, while recall decreased. The F1-score peaked when confidence threshold is 80\% (F1=0.883), demonstrating that our qualitatively-grounded threshold is also empirically suitable for balancing precision and recall.

\section{Analysis on Temporal Trends}\label{app:temporal}

We conducted an analysis on the temporal trends of \proj{}'s annotation to explore whether conformity effects emerged. \textbf{We first found the annotation size becomes more granular with the annotation progression, suggesting that participants are probably meticulously examining the annotation.} Annotation size was measured through the bounding box area relative to the screen size, consistent with Section~\ref{sec:distribution}. RM-ANOVA\footnote{Prior to analysis, Shapiro-Wilk tests confirmed that data met the normality assumption necessary for parametric testing.} revealed a significant main effect of Day on bounding box area ($F_{6, 174} = 43.5$, $p < .001$). Post-hoc Tukey HSD test found a significant difference between Day 1 and Day 7 ($p < .05$), showing that users moved from marking broad, generalized regions to identify precise and localized artifacts. This contrasts to conformity bias, where one might stabilize the generic annotation size without refinement. 

\textbf{We then found the semantic distribution of labels also exhibited an evolution over time.} We performed a Chi-Square test of independence to assess the relationship between the time period and the distribution of label categories\footnote{Due to the non-parametric nature, we adopted the non-parametric tests.}. The analysis yielded a significant result ($\chi^2 = 96.1$, $p < .001$), indicating that the composition of label types shifted significantly. Specifically, the proportion of the generic `mismatch' declined, while specific, evidence-based categories such as `blurry' and `unnatural skin' increased. This suggests a qualitative transformation in user behavior, where the vocabulary used to describe deepfake artifacts became increasingly specific and diagnostic. This contrasts to conformity bias, where users may converge toward the initial, low-effort generic labels. 

\textbf{We finally found an increasing degree of spatio-temporal overlap between individual user dimensions and the aggregated consensus.} We quantified this using the 3D Intersection-over-Union (IoU) metric, which calculates the volumetric overlap of annotated regions across both spatial and temporal dimensions. A RM-ANOVA\footnote{Normality assumptions were verified prior to conducting the RM-ANOVA.} revealed a highly significant main effect of Day on IoU scores ($F_{6,174} = 50.3$, $p < .001$). The average IoU score demonstrated a upward trend, rising from approximately 0.05 on Day 1 to 0.75 on Day 7. Post-hoc analysis using Tukey's HSD tests confirmed that this difference between the initial and final days was statistically significant ($p < .05$). 

This convergence indicates that users increasingly focused on the same regions over time. We interpret this as a constructive consensus rather than blind conformity. If users were simply mimicking peer inputs, we would expect to see high overlap accompanied by static bounding box sizes and generic labels. However, our findings reveal a different pattern when combining the three aspects: (1) High overlap: users looked at the same location, (2) High precision: users actively shrank the bounding boxes to be more granular, (3) Semantic shift: users replaced generic labels with more specific labels. Therefore, the aggregated annotations likely served as a facilitator in the users' annotation process. While we acknowledge that users do exhibit conformity behavior, we interpret that as a constructive signal, where users engaged in verification to refine the boundaries and descriptions rather than passive compliance. 

Finally, we advocate for future research to systematically analyze the conformity effects of spatio-temporal annotations using a stricter protocol. This protocol should clearly distinguish these effects through experimental design~\cite{cherng2022understanding,bauer2020conformity}. This goal could potentially be achieved through adding groups with (1) an alternative label demonstration mechanism, where users first annotate by themselves, and are only shown to peer labels after their initial annotation, or (2) customized toggles, where users could choose when and how to view others' annotations. 



\end{document}